\catcode`\@=11

\font\tenmsa=msam10
\font\sevenmsa=msam7
\font\fivemsa=msam5
\font\tenmsb=msbm10
\font\sevenmsb=msbm7
\font\fivemsb=msbm5
\newfam\msafam
\newfam\msbfam
\textfont\msafam=\tenmsa  \scriptfont\msafam=\sevenmsa
  \scriptscriptfont\msafam=\fivemsa
\textfont\msbfam=\tenmsb  \scriptfont\msbfam=\sevenmsb
  \scriptscriptfont\msbfam=\fivemsb

\def\hexnumber@#1{\ifcase#1 0\or1\or2\or3\or4\or5\or6\or7\or8\or9\or
A\or B\or C\or D\or E\or F\fi }

\font\teneuf=eufm10
\font\seveneuf=eufm7
\font\fiveeuf=eufm5
\newfam\euffam
\textfont\euffam=\teneuf
\scriptfont\euffam=\seveneuf
\scriptscriptfont\euffam=\fiveeuf
\def\frak{\ifmmode\let\next\frak@\else
 \def\next{\Err@{Use \string\frak\space only in math mode}}\fi\next}
\def\goth{\ifmmode\let\next\frak@\else
 \def\next{\Err@{Use \string\goth\space only in math mode}}\fi\next}
\def\frak@#1{{\frak@@{#1}}}
\def\frak@@#1{\fam\euffam#1}

\edef\msa@{\hexnumber@\msafam}
\edef\msb@{\hexnumber@\msbfam}

\mathchardef\boxdot="2\msa@00
\mathchardef\boxplus="2\msa@01
\mathchardef\boxtimes="2\msa@02
\mathchardef\square="0\msa@03
\mathchardef\blacksquare="0\msa@04
\mathchardef\centerdot="2\msa@05
\mathchardef\lozenge="0\msa@06
\mathchardef\blacklozenge="0\msa@07
\mathchardef\circlearrowright="3\msa@08
\mathchardef\circlearrowleft="3\msa@09
\mathchardef\rightleftharpoons="3\msa@0A
\mathchardef\leftrightharpoons="3\msa@0B
\mathchardef\boxminus="2\msa@0C
\mathchardef\Vdash="3\msa@0D
\mathchardef\Vvdash="3\msa@0E
\mathchardef\vDash="3\msa@0F
\mathchardef\twoheadrightarrow="3\msa@10
\mathchardef\twoheadleftarrow="3\msa@11
\mathchardef\leftleftarrows="3\msa@12
\mathchardef\rightrightarrows="3\msa@13
\mathchardef\upuparrows="3\msa@14
\mathchardef\downdownarrows="3\msa@15
\mathchardef\upharpoonright="3\msa@16

\mathchardef\downharpoonright="3\msa@17
\mathchardef\upharpoonleft="3\msa@18
\mathchardef\downharpoonleft="3\msa@19
\mathchardef\rightarrowtail="3\msa@1A
\mathchardef\leftarrowtail="3\msa@1B
\mathchardef\leftrightarrows="3\msa@1C
\mathchardef\rightleftarrows="3\msa@1D
\mathchardef\Lsh="3\msa@1E
\mathchardef\Rsh="3\msa@1F
\mathchardef\rightsquigarrow="3\msa@20
\mathchardef\leftrightsquigarrow="3\msa@21
\mathchardef\looparrowleft="3\msa@22
\mathchardef\looparrowright="3\msa@23
\mathchardef\circeq="3\msa@24
\mathchardef\succsim="3\msa@25
\mathchardef\gtrsim="3\msa@26
\mathchardef\gtrapprox="3\msa@27
\mathchardef\multimap="3\msa@28
\mathchardef\therefore="3\msa@29
\mathchardef\because="3\msa@2A
\mathchardef\doteqdot="3\msa@2B

\mathchardef\triangleq="3\msa@2C
\mathchardef\precsim="3\msa@2D
\mathchardef\lesssim="3\msa@2E
\mathchardef\lessapprox="3\msa@2F
\mathchardef\eqslantless="3\msa@30
\mathchardef\eqslantgtr="3\msa@31
\mathchardef\curlyeqprec="3\msa@32
\mathchardef\curlyeqsucc="3\msa@33
\mathchardef\preccurlyeq="3\msa@34
\mathchardef\leqq="3\msa@35
\mathchardef\leqslant="3\msa@36
\mathchardef\lessgtr="3\msa@37
\mathchardef\backprime="0\msa@38
\mathchardef\risingdotseq="3\msa@3A
\mathchardef\fallingdotseq="3\msa@3B
\mathchardef\succcurlyeq="3\msa@3C
\mathchardef\geqq="3\msa@3D
\mathchardef\geqslant="3\msa@3E
\mathchardef\gtrless="3\msa@3F
\mathchardef\sqsubset="3\msa@40
\mathchardef\sqsupset="3\msa@41
\mathchardef\vartriangleright="3\msa@42
\mathchardef\vartriangleleft="3\msa@43
\mathchardef\trianglerighteq="3\msa@44
\mathchardef\trianglelefteq="3\msa@45
\mathchardef\bigstar="0\msa@46
\mathchardef\between="3\msa@47
\mathchardef\blacktriangledown="0\msa@48
\mathchardef\blacktriangleright="3\msa@49
\mathchardef\blacktriangleleft="3\msa@4A
\mathchardef\vartriangle="0\msa@4D
\mathchardef\blacktriangle="0\msa@4E
\mathchardef\triangledown="0\msa@4F
\mathchardef\eqcirc="3\msa@50
\mathchardef\lesseqgtr="3\msa@51
\mathchardef\gtreqless="3\msa@52
\mathchardef\lesseqqgtr="3\msa@53
\mathchardef\gtreqqless="3\msa@54
\mathchardef\Rrightarrow="3\msa@56
\mathchardef\Lleftarrow="3\msa@57
\mathchardef\veebar="2\msa@59
\mathchardef\barwedge="2\msa@5A
\mathchardef\doublebarwedge="2\msa@5B
\mathchardef\angle="0\msa@5C
\mathchardef\measuredangle="0\msa@5D
\mathchardef\sphericalangle="0\msa@5E
\mathchardef\varpropto="3\msa@5F
\mathchardef\smallsmile="3\msa@60
\mathchardef\smallfrown="3\msa@61
\mathchardef\Subset="3\msa@62
\mathchardef\Supset="3\msa@63
\mathchardef\Cup="2\msa@64

\mathchardef\Cap="2\msa@65

\mathchardef\curlywedge="2\msa@66
\mathchardef\curlyvee="2\msa@67
\mathchardef\leftthreetimes="2\msa@68
\mathchardef\rightthreetimes="2\msa@69
\mathchardef\subseteqq="3\msa@6A
\mathchardef\supseteqq="3\msa@6B
\mathchardef\bumpeq="3\msa@6C
\mathchardef\Bumpeq="3\msa@6D
\mathchardef\lll="3\msa@6E

\mathchardef\ggg="3\msa@6F

\mathchardef\circledS="0\msa@73
\mathchardef\pitchfork="3\msa@74
\mathchardef\dotplus="2\msa@75
\mathchardef\backsim="3\msa@76
\mathchardef\backsimeq="3\msa@77
\mathchardef\complement="0\msa@7B
\mathchardef\intercal="2\msa@7C
\mathchardef\circledcirc="2\msa@7D
\mathchardef\circledast="2\msa@7E
\mathchardef\circleddash="2\msa@7F
\def\ulcorner{\delimiter"4\msa@70\msa@70 }
\def\urcorner{\delimiter"5\msa@71\msa@71 }
\def\llcorner{\delimiter"4\msa@78\msa@78 }
\def\lrcorner{\delimiter"5\msa@79\msa@79 }
\def\yen{\mathhexbox\msa@55 }
\def\checkmark{\mathhexbox\msa@58 }
\def\circledR{\mathhexbox\msa@72 }
\def\maltese{\mathhexbox\msa@7A }
\mathchardef\lvertneqq="3\msb@00
\mathchardef\gvertneqq="3\msb@01
\mathchardef\nleq="3\msb@02
\mathchardef\ngeq="3\msb@03
\mathchardef\nless="3\msb@04
\mathchardef\ngtr="3\msb@05
\mathchardef\nprec="3\msb@06
\mathchardef\nsucc="3\msb@07
\mathchardef\lneqq="3\msb@08
\mathchardef\gneqq="3\msb@09
\mathchardef\nleqslant="3\msb@0A
\mathchardef\ngeqslant="3\msb@0B
\mathchardef\lneq="3\msb@0C
\mathchardef\gneq="3\msb@0D
\mathchardef\npreceq="3\msb@0E
\mathchardef\nsucceq="3\msb@0F
\mathchardef\precnsim="3\msb@10
\mathchardef\succnsim="3\msb@11
\mathchardef\lnsim="3\msb@12
\mathchardef\gnsim="3\msb@13
\mathchardef\nleqq="3\msb@14
\mathchardef\ngeqq="3\msb@15
\mathchardef\precneqq="3\msb@16
\mathchardef\succneqq="3\msb@17
\mathchardef\precnapprox="3\msb@18
\mathchardef\succnapprox="3\msb@19
\mathchardef\lnapprox="3\msb@1A
\mathchardef\gnapprox="3\msb@1B
\mathchardef\nsim="3\msb@1C
\mathchardef\ncong="3\msb@1D

\mathchardef\varsubsetneq="3\msb@20
\mathchardef\varsupsetneq="3\msb@21
\mathchardef\nsubseteqq="3\msb@22
\mathchardef\nsupseteqq="3\msb@23
\mathchardef\subsetneqq="3\msb@24
\mathchardef\supsetneqq="3\msb@25
\mathchardef\varsubsetneqq="3\msb@26
\mathchardef\varsupsetneqq="3\msb@27
\mathchardef\subsetneq="3\msb@28
\mathchardef\supsetneq="3\msb@29
\mathchardef\nsubseteq="3\msb@2A
\mathchardef\nsupseteq="3\msb@2B
\mathchardef\nparallel="3\msb@2C
\mathchardef\nmid="3\msb@2D
\mathchardef\nshortmid="3\msb@2E
\mathchardef\nshortparallel="3\msb@2F
\mathchardef\nvdash="3\msb@30
\mathchardef\nVdash="3\msb@31
\mathchardef\nvDash="3\msb@32
\mathchardef\nVDash="3\msb@33
\mathchardef\ntrianglerighteq="3\msb@34
\mathchardef\ntrianglelefteq="3\msb@35
\mathchardef\ntriangleleft="3\msb@36
\mathchardef\ntriangleright="3\msb@37
\mathchardef\nleftarrow="3\msb@38
\mathchardef\nrightarrow="3\msb@39
\mathchardef\nLeftarrow="3\msb@3A
\mathchardef\nRightarrow="3\msb@3B
\mathchardef\nLeftrightarrow="3\msb@3C
\mathchardef\nleftrightarrow="3\msb@3D
\mathchardef\divideontimes="2\msb@3E
\mathchardef\varnothing="0\msb@3F
\mathchardef\nexists="0\msb@40
\mathchardef\mho="0\msb@66
\mathchardef\eth="0\msb@67
\mathchardef\eqsim="3\msb@68
\mathchardef\beth="0\msb@69
\mathchardef\gimel="0\msb@6A
\mathchardef\daleth="0\msb@6B
\mathchardef\lessdot="3\msb@6C
\mathchardef\gtrdot="3\msb@6D
\mathchardef\ltimes="2\msb@6E
\mathchardef\rtimes="2\msb@6F
\mathchardef\shortmid="3\msb@70
\mathchardef\shortparallel="3\msb@71
\mathchardef\smallsetminus="2\msb@72
\mathchardef\thicksim="3\msb@73
\mathchardef\thickapprox="3\msb@74
\mathchardef\approxeq="3\msb@75
\mathchardef\succapprox="3\msb@76
\mathchardef\precapprox="3\msb@77
\mathchardef\curvearrowleft="3\msb@78
\mathchardef\curvearrowright="3\msb@79
\mathchardef\digamma="0\msb@7A
\mathchardef\varkappa="0\msb@7B
\mathchardef\hslash="0\msb@7D
\mathchardef\hbar="0\msb@7E
\mathchardef\backepsilon="3\msb@7F
\def\Bbb{\ifmmode\let\next\Bbb@\else
 \def\next{\errmessage{Use \string\Bbb\space only in math mode}}\fi\next}
\def\Bbb@#1{{\Bbb@@{#1}}}
\def\Bbb@@#1{\fam\msbfam#1}

\catcode`\@=12

\def\m@th{\mathsurround=0pt}
\newif\ifdtpt
\def\displ@y{\openup1\jot\m@th
    \everycr{\noalign{\ifdtpt\dt@pfalse
    \vskip-\lineskiplimit \vskip\normallineskiplimit
    \else \penalty\interdisplaylinepenalty \fi}}}
%
\def\eqalignc#1{\,\vcenter{\openup1\jot\m@th
                \ialign{\strut\hfil$\displaystyle{##}$\hfil&
                              \hfil$\displaystyle{{}##}$\hfil&
                              \hfil$\displaystyle{{}##}$\hfil&
                              \hfil$\displaystyle{{}##}$\hfil&
                              \hfil$\displaystyle{{}##}$\hfil\crcr#1\crcr}}\,}
\def\eqalignnoc#1{\displ@y\tabskip\centering \halign to \displaywidth{
                  \hfil$\displaystyle{##}$\hfil\tabskip=0pt &
                  \hfil$\displaystyle{{}##}$\hfil\tabskip=0pt &
                  \hfil$\displaystyle{{}##}$\hfil\tabskip=0pt &
                  \hfil$\displaystyle{{}##}$\hfil\tabskip=0pt &
                  \hfil$\displaystyle{{}##}$\hfil\tabskip\centering &
                  \llap{$##$}\tabskip=0pt \crcr#1\crcr}}
\def\leqalignnoc#1{\displ@y\tabskip\centering \halign to \displaywidth{
                  \hfil$\displaystyle{##}$\hfil\tabskip=0pt &
                  \hfil$\displaystyle{{}##}$\hfil\tabskip=0pt &
                  \hfil$\displaystyle{{}##}$\hfil\tabskip=0pt &
                  \hfil$\displaystyle{{}##}$\hfil\tabskip=0pt &
                  \hfil$\displaystyle{{}##}$\hfil\tabskip\centering &
                  \kern-\displaywidth\rlap{$##$}\tabskip=\displaywidth
                  \crcr#1\crcr}}
%
\def\dasharrowfill{$\mathsurround=0pt \mathord- \mkern-6mu
    \cleaders\hbox{$\mkern-2mu\mathord-\mkern-2mu$}\hfill
    \mkern-6mu \mathord-$}

%


\def\charlvmidlw#1#2{\,\vtop{\ialign{##\crcr
      #1\crcr\noalign{\kern1pt\nointerlineskip}
      $\hfil#2\hfil$\crcr}}\,}
\def\charlvlowlw#1#2{\,\vtop{\ialign{##\crcr
      $\hfil#1\hfil$\crcr\noalign{\kern1pt\nointerlineskip}
      #2\crcr}}\,}
\def\charlvmidup#1#2{\,\vbox{\ialign{##\crcr
      $\hfil#1\hfil$\crcr\noalign{\kern1pt\nointerlineskip}
      #2\crcr}}\,}
\def\charlvupup#1#2{\,\vbox{\ialign{##\crcr
      #1\crcr\noalign{\kern1pt\nointerlineskip}
      $\hfil#2\hfil$\crcr}}\,}

\def\vspce{\kern4pt} \def\hspce{\kern4pt}    

\def\emptybox{\vbox{\kern.7ex\hbox{\kern.5em}\kern.7ex}}
 \font\sevmi  = cmmi7              
    \skewchar\sevmi ='177
 \font\fivmi  = cmmi5              
    \skewchar\fivmi ='177
\font\tenmib=cmmib10
\newfam\bfmitfam

\textfont\bfmitfam=\tenmib
\scriptfont\bfmitfam=\sevmi
\scriptscriptfont\bfmitfam=\fivmi
%

%

\mathchardef\gq="0060
\mathchardef\dq="0027
\mathchardef\ssmath="19
\mathchardef\aemath="1A
\mathchardef\oemath="1B
\mathchardef\omath="1C
\mathchardef\AEmath="1D
\mathchardef\OEmath="1E
\mathchardef\Omath="1F
\mathchardef\imath="10
\mathchardef\fmath="0166
\mathchardef\gmath="0167
\mathchardef\vmath="0176

%

\def\dotplus{\buildrel\textstyle .\over +}

\def\twodot{.\kern-0.1em.}

\def\paral{\mathrel{/\kern-.25em/}}
\def\grlo{\mathrel{\hbox{\lower.2ex\hbox{\rlap{$>$}\raise1ex\hbox{$<$}}}}}
\def\logr{\mathrel{\hbox{\lower.2ex\hbox{\rlap{$<$}\raise1ex\hbox{$>$}}}}}
\def\greq{\mathrel{\hbox{\lower1ex\hbox{\rlap{$=$}\raise1.2ex\hbox{$>$}}}}}
\def\loeq{\mathrel{\hbox{\lower1ex\hbox{\rlap{$=$}\raise1.2ex\hbox{$<$}}}}}
\def\grsim{\mathrel{\hbox{\lower1ex\hbox{\rlap{$\sim$}\raise1ex\hbox{$>$}}}}}
\def\losim{\mathrel{\hbox{\lower1ex\hbox{\rlap{$\sim$}\raise1ex\hbox{$<$}}}}}
%
\font\ninerm=cmr9
\def\uniset{\rlap{\ninerm 1}\kern.15em 1}

\def\emptysq{\mathbin{\vbox{\hrule\hbox{\vrule height1ex \kern.5em
                            \vrule height1ex}\hrule}}}
\def\emptyrect{\mathbin{\vbox{\hrule\hbox{\vrule height1ex \kern1em
                              \vrule height1ex}\hrule}}}
\def\rightonleftarrow{\mathrel{\hbox{\raise.5ex\hbox{$\rightarrow$}\ignorespaces
                                   \lower.5ex\hbox{\llap{$\leftarrow$}}}}}
\def\leftonrightarrow{\mathrel{\hbox{\raise.5ex\hbox{$\leftarrow$}\ignorespaces
                                   \lower.5ex\hbox{\llap{$\rightarrow$}}}}}

\def\bkB{{\rm I\kern-.17em B}}
\def\bkC{{\rm \kern.24em
            \vrule width.05em height1.4ex depth-.05ex
            \kern-.26em C}}
\def\bkD{{\rm I\kern-.17em D}}
\def\bkE{{\rm I\kern-.17em E}}
\def\bkF{{\rm I\kern-.17em F}}
\def\bkG{{\rm \kern.24em
            \vrule width.05em height1.4ex depth-.05ex
            \kern-.26em G}}
\def\bkH{{\rm I\kern-.22em H}}
\def\bkI{{\rm I\kern-.22em I}}
\def\bkJ{{\rm \kern.19em
            \vrule width.02em height1.5ex depth0ex
            \kern-.20em J}}
\def\bkK{{\rm I\kern-.22em K}}
\def\bkL{{\rm I\kern-.17em L}}
\def\bkM{{\rm I\kern-.22em M}}
\def\bkN{{\rm I\kern-.20em N}}
\def\bkO{{\rm \kern.24em
            \vrule width.05em height1.4ex depth-.05ex
            \kern-.26em O}}
\def\bkP{{\rm I\kern-.17em P}}
\def\bkQ{{\rm \kern.24em
            \vrule width.05em height1.4ex depth-.05ex
            \kern-.26em Q}}
\def\bkR{{\rm I\kern-.17em R}}
\def\bkT{{\rm \kern.24em
            \vrule width.02em height1.5ex depth 0ex
            \kern-.27em T}}
\def\bkU{{\rm \kern.30em
            \vrule width.02em height1.47ex depth-.05ex
            \kern-.32em U}}
\def\bkZ{{\rm Z\kern-.32em Z}}
%

\centerline{{\bf TOWARDS A RELATIVISTIC KMS-CONDITION}}

\vglue 1truecm

\centerline{{\bf Jacques Bros}}
\centerline{Service de Physique Th\'eorique}
\centerline{Laboratoire de la Direction des Sciences de la Mati\`ere}
\centerline{CE-Saclay, 91191 Gif-sur-Yvette Cedex, FRANCE}
\centerline{{\bf Detlev Buchholz}}
\centerline{II. Institut f\"ur Theoretische Physik}
\centerline{Universit\"at Hamburg, D-22761 Hamburg, GERMANY}

\vglue 2truecm

\centerline{ABSTRACT}
\vglue 0.5truecm

\noindent It is shown that, under quite general conditions, thermal
correlation functions in relativistic quantum field theory have stronger
analyticity properties in configuration space than those imposed by the
KMS-condition. These analyticity properties may be understood as a
remnant of
the relativistic spectrum condition in the vacuum sector and lead to a
Lorentz-covariant formulation of the KMS-condition involving all
space-time variables.

\vfill\eject

\noindent {\bf 1. Introduction}

\vskip 12pt

\noindent In relativistic quantum field theory, one characterizes
thermal equilibrium
states in the same way as their counterparts in non-relativistic
quantum
statistical mechanics by means of the KMS-condition
$\lbrack$BR,H$\rbrack$. If
one denotes by $ {\cal A} $ the
algebra of observables and by
$ \alpha_ t $ the automorphism of $ {\cal A} $
inducing the time
translations in the rest frame of the considered heat bath,
this condition
reads as follows\nobreak\ :

\vskip 12pt

\noindent {\sl KMS-Condition\/}\nobreak\ :
{\it The state $ \omega_ \beta $ on
$ {\cal A} $ satisfies the KMS-condition at inverse
temperature $ \beta  > 0 $ iff for every pair of operators
$ A,B \in  {\cal A} $ there exists an
analytic function $ F $ in the strip
$ S_\beta  = \{ z\in  \Bbb C : 0 < {\rm Im} \ z < \beta\} $ with
continuous boundary values at $ {\rm Im} \ z = 0 $ and
$ {\rm Im} \ z = i\beta, $ given respectively (for
$ t\in  \Bbb R) $ by:
$$ F(t) = \omega_ \beta \left(A\alpha_ t(B) \right),\ \ \ \ \
F(t+i \beta )  = \omega_ \beta \left(\alpha_ t(B)A \right)\ . $$ }

The notions used in this quite general formulation of the condition
are related to the conventional field-theoretical setting
as follows. The algebra ${\cal A}$ may be thought of as
being generated by bounded functions of the underlying observable
fields, currents, etc. If $\phi ( x )$ is any such field and if $f ( x
) $ is any real test function with support in a bounded region of
spacetime, then the corresponding (unitary)
operator $ A = e^{ i \int dx \, f(x) \phi ( x )} $ would be a typical
element of ${\cal A}$. Conversely, one can recover the fields from such
operators in ${\cal A}$ by taking (functional) derivatives. We assume
here that the algebra ${\cal A}$ is defined on the vacuum Hilbert space
${\cal H}$ of the theory. The vacuum state $\omega_\infty$ can then be
represented in the form $ \omega_\infty ( A ) = ( \Omega, A \Omega ),
\, A \in {\cal A}$, where $\Omega \in {\cal H}$ is the vacuum vector.
Similarly, the thermal states $\omega_\beta$ are positive, linear
and normalized functionals on ${\cal A}$ which, according to the
reconstruction theorem, can be represented by vectors $\Omega_\beta$
in the thermal Hilbert spaces ${\cal H}_\beta $. Thus the functions
$F ( t )$ in the formulation of the KMS-condition
are the correlation functions of an arbitrary pair of
observables in these states. The automorphism $\alpha_t$ inducing
the time translations can be represented on the vacuum Hilbert space
${\cal H}$ according to
$\alpha_t ( A ) = e^{ i t H } A e^{ - i t H }, \,
A \in {\cal A}$, where $H$ is the Hamiltonian. A similar representation
holds on the thermal Hilbert spaces ${\cal H}_\beta$. But, whereas the
action of
$\alpha_t$ on ${\cal A}$ does not depend on the state $\omega_\beta$
which one considers, the respective unitaries $ e^{i t H_\beta}$
implementing this action on the spaces ${\cal H}_\beta$ do.
The present somewhat abstract
approach is therefore appropriate if one wants to consider
all thermal states at the same time.

The characterization of equilibrium states by the KMS-condition
recalled above is well
justified $\lbrack$HHW, HKTP, PW$\rbrack$, and the breaking of Lorentz
invariance in this
condition through the choice of a distinguished time-axis is fully
understood $\lbrack$O,N$\rbrack$. Nevertheless it would seem natural in
a relativistic setting to incorporate in the KMS-condition the
properties of equilibrium states with
respect to observations {\sl in arbitrary Lorentz frames\/}.
It is the aim of the present article to shed some light on this problem.

Let us begin with some considerations based on a
property of thermal equilibrium states about which all
uniformly moving observers ought to agree\nobreak\ : it is impossible to
extract from
such a state an arbitrarily large amount of energy by local operations,
namely the energy content of equilibrium states is locally finite. We
first advocate that, for an observer in the rest frame of the state,
this property is essentially encoded in the analyticity
requirement of the KMS-condition.

In fact, let us make the assumption considered as
physically well-motivated in the general algebraic setting of quantum
field theory, that the KMS-state $ \omega_ \beta $ is locally normal
with respect to the vacuum sector $\lbrack$H$\rbrack$. By
applying standard arguments in the theory of operator algebras
$\lbrack$D$\rbrack$, one deduces from this
assumption that for each bounded space-time region $ {\cal O} $ there
exists some vector $ \Omega_{ \beta ,{\cal O}}\in{\cal H}, $ where $
{\cal H} $ is the Hilbert space describing the vacuum sector, such that
$ \omega_ \beta( A)= \left(\Omega_{ \beta ,{\cal O}},A\ \Omega_{ \beta
,{\cal O}} \right) $
for all observables $ A $ which are localized in $ {\cal O}. $ The
KMS-condition for $ \omega_ \beta $ then entails that for observables $
A $ which are localized in the interior of $ {\cal O} $ and
for sufficiently small $ \vert t\vert $ the vector valued function $ t
\longrightarrow \alpha_ t(A)\Omega_{ \beta ,{\cal O}} $ has an analytic
continuation into the strip $ S_{\beta /2}. $ Hence $ \Omega_{ \beta
,{\cal O}} $ behaves \lq\lq locally\rq\rq\ in the same way as an
analytic vector for the energy $ H $ (lying in the domain
of $ {\rm e}^{{\beta \over 2}H}). $ This indicates that large local
energy contributions, though present due to energy fluctuations, are
exponentially suppressed in the state $ \omega_ \beta. $

Since in the vacuum sector of a relativistic theory the notion of \lq\lq
analytic vector for the energy\rq\rq\ is a Lorentz invariant concept
(as a consequence
of the relativistic spectrum condition), the above argument suggests
that, with
respect to all time-like directions $ f \in  V_+, $ the function $ t
\longrightarrow  \alpha^{( f)}_t (A)\Omega_{ \beta ,{\cal O}} $
should have, for sufficiently small $ \vert t\vert , $ an analytic
continuation into some domain of the upper half plane. Because of local
normality this would in turn imply that also the correlation functions
$ t \longrightarrow  \omega_ \beta \left(A\ \alpha^{( f)}_t(B) \right)
$ can be continued
analytically into that domain. This condition could be regarded as a
{\sl stability requirement\/}, giving a formal
expression to the idea that the local
energy content of equilibrium states is finite in all Lorentz frames.

Unfortunately we have not been able to cast these heuristic
considerations into a rigorous argument without further input, the
subtle point being the question in which precise sense the vectors $
\Omega_{ \beta ,{\cal O}} $ may be regarded as being
locally analytic for the energy operator in the various Lorentz frames.

We will therefore adopt a different approach which relies in
its spirit on the original introduction of KMS states (due to
$\lbrack$HHW$\rbrack$) as thermodynamic limits of appropriate local
Gibbs states. The method which we will apply for carrying out this
approach in a rigorous way is the one introduced in
$\lbrack$BJ$\rbrack$ whose main result is the following: if the
underlying theory satisfies an
appropriate \lq\lq nuclearity condition\rq\rq\ proposed in
$\lbrack$BW$\rbrack$ which restricts the number of local degrees of
freedom in a physically sensible manner, then, in a generic way,
KMS states $ \omega_\beta$ can be approximated
by states representable by a rigorous (local) version of the
Gibbs formula $ Z^{-1} {\rm e}^{-\beta H} $.
For the correlation functions of these
approximations, the relativistic
spectrum condition can be applied and it has implications in terms of
space-time analyticity
properties which are of the type indicated above.
Then, if the thermodynamic limit can be controlled
sufficiently well, namely if long range boundary effects
are negligible in a way which will be made precise below, we can
establish similar analyticity properties of the correlation functions.

These analyticity properties are a remnant of the spectral properties of
energy and momentum in the vacuum sector and may be regarded as an
appropriate substitute to the relativistic spectrum
condition for the case of thermal equilibrium states.
Moreover, they suggest a specific Lorentz-covariant
formulation of the KMS-condition. In the simplest case of equilibrium
states in a minkowskian
space $ \Bbb R^d $ which are invariant
under space-time translations and unsensitive to boundary effects, this
condition reads as follows $ (V_+ $ and $ V_- $ denoting respectively
the
cones of future
and past events in $ \Bbb R^d): $

\vskip 12pt

\noindent {\sl Relativistic KMS-condition\/}\nobreak\ : {\it The state $
\omega_ \beta $ on $ {\cal A} $ satisfies the
relativistic KMS-condition at inverse temperature $ \beta  > 0 $ iff
there
exists some
positive timelike vector $ e\in V_+, $ $ e^2=1, $ such that for every
pair of
local
operators $ A,B \in  {\cal A} $ there is a function $ F $ which is
analytic in
the tube $ {\cal T}_{\beta e}= \left\{ z\in  \Bbb C^d: {\rm Im} \ z\in
V_+\cap
\left(\beta e+V_- \right) \right\} $
and continuous at the boundary sets\footnote{$ ^{1)} $}{{\sevenrm The
precise
definition
of continuity at the edges of tubes with conical bases is given in
Appendix
A.}} $ {\rm Im} \ z=0, $ $ {\rm Im} \ z = \beta e $ with boundary
values given
by
$$ F(x) = \omega_ \beta \left(A\alpha_ x(B) \right),\ \ \ \ F(x+i \beta
e) =
\omega_ \beta \left(\alpha_ x(B)A \right) $$
for $ x\in  \Bbb R^d. $
}

In this condition, $ \alpha_ x $ denotes the automorphism of $ {\cal A}
$
which induces the
space-time translations in $ \Bbb R^d, $ all space-time variables $
x\in \Bbb
R^d $ being treated on an equal
footing. The condition thereby applies to all observers moving with
constant
velocity and allows them to identify $ \omega_ \beta $ as an
equilibrium state
which fixes
a distinguished rest frame with time direction along $ e. $

It may be instructive to illustrate this condition on the case where
$A, B$ are replaced respectively by field operators
$  \phi (f) = \int dx \, f(x) \phi ( x ) $ and
$  \phi (g) = \int dx \, g(x) \phi ( x ) $
so that
$\omega_\beta ( \phi ( f ) \phi ( g ) ) = \int dx \int dy
f ( x ) g ( y ) {\cal W}_\beta ( x - y ) $, where
$ {\cal W}_\beta ( x - y )$ is the thermal two-point (Wightman)
function of the field $\phi$. The relativistic
KMS-condition then amounts to the following analyticity property
of ${\cal W}_\beta$: There exists a function $W_\beta$, analytic
in the tube ${\cal T}_{\beta e}$, with boundary values given by
$W_\beta ( x ) =  {\cal W}_\beta (-x)$ and $ W_\beta ( x + i \beta e )
=  {\cal W}_\beta ( x )$. Analogous analyticity properties can be
stated for the $n$-point functions.

We believe that the relativistic KMS-condition covers a large area of
equilibrium situations with the
possible
exception of phase-transition points, where boundary effects matter.
Some
variants of the condition, dealing with equilibrium states which are
more
sensitive to boundary effects or not invariant under (space)
translations
will be derived in the main text.

In the subsequent Sec.\ 2 we state our assumptions and recall the
construction
of thermal equilibrium states given in $\lbrack$BJ$\rbrack$, which is
based on
an
approximating family of local equilibrium states in the vacuum sector.
We
will then exhibit (in Sec.\ 3) analyticity properties of these
approximations
with respect to space-time translations which follow from the
relativistic
spectrum condition. Sec.\ 4 is devoted to a study of the influence of
boundary
effects on the analyticity properties of the limit states. These
results are
taken as an input in Sec.\ 5 for determining the primitive domains of
analyticity of thermal correlation functions in various cases\nobreak\
; the
pertinent
mathematical facts used in this analysis are gathered in two
appendices. It
turns out that the correlation functions exhibit analyticity properties
with
respect to the space-time variables which corroborate in a rigorous way
the
conclusions of the previous heuristic discussion. The article concludes
with
a remark on a possible alternative approach towards the justification
of a
relativistic KMS-condition.

\vskip 24pt

\noindent {\bf 2. Local approximations of KMS-states}

\vskip 12pt

\noindent For analysing the consequences of the
relativistic spectrum condition for the
structure of KMS-states, it is necessary to know how these states are related
to the vacuum sector. Such a link has been established in $\lbrack$BJ$\rbrack$
for theories
with a reasonable number of local degrees of freedom. Since our
argument is
based on that approach, we recall here the relevant assumptions and
results of
$\lbrack$BJ$\rbrack$.

The setting used in this investigation is algebraic quantum field
theory,
where basic physical principles such as locality, relativistic
covariance and
stability of matter are expressed in terms of a family of algebras $
{\cal
A}({\cal O}) $
representing the observables of the underlying theory which are
localized in
the space-time region $ {\cal O}, $ (cf.$\lbrack$H$\rbrack$).

\vskip 12pt
1.{\sl\ (Locality)\/} There is a family of (concrete) $ C^\ast
$-algebras $
{\cal A}({\cal O}) $ which are
labelled by the open bounded space-time regions $ {\cal O}\subset \Bbb
R^d $
and act on a Hilbert
space $ {\cal H} $ representing the states in the vacuum sector. These
algebras are
subject to the condition of isotony,
$$ {\cal A} \left({\cal O}_1 \right)\subset{\cal A} \left({\cal O}_2
\right)\
\ \ {\rm if} \ \ \ {\cal O}_1 \subset  {\cal O}_2\ ; \eqno (2.1) $$
so the assignment $ {\cal O} \longrightarrow{\cal A}({\cal O}) $
defines a
local net over $ \Bbb R^d. $ The $ C^\ast $-inductive
limit of this net is denoted by $ {\cal A} $ and assumed to act
irreducibly on
$ {\cal H}. $ (We
note that in this investigation we do not rely on any form of spacelike
commutation relations.)

\vskip 12pt

2.{\sl\ (Covariance)\/} The space-time translations $ x\in \Bbb R^d $
act on $
{\cal A} $ by
automorphisms $ \alpha_ x $ which are unitarily implemented on $ {\cal
H} $ by
$$ \alpha_ x(A) = {\rm e}^{iPx} A\ {\rm e}^{-iPx},\ \ \ \  A \in {\cal
A} \eqno (2.2) $$
where $ P $ are the energy-momentum operators. The action of $ \alpha_
x $ on
$ {\cal A} $ is strongly
continuous\footnote{$ ^{2)} $}{{\sevenrm More specifically, there holds
$
\lim_{x \longrightarrow 0} \left\Vert \alpha_ x(A)-A \right\Vert =0 $
for all $ A\in{\cal A}, $ where $ \Vert \cdot\Vert $ is the operator
norm.}}
and covariant, i.e.
$$ \alpha_ x({\cal A}({\cal O})) = {\cal A}({\cal O}+x)\ ,
\eqno (2.3) $$
where $ {\cal O}+x $ is the region $ {\cal O} $ shifted by $ x. $

\vskip 12pt

3.{\sl\ (Stability)\/} The energy-momentum operators $ P $ satisfy the
relativistic
spectrum condition, $ {\rm sp} \ P \subset  \bar V_+, $ and there is a
unique
ground state $ \Omega  \in  {\cal H}, $ such
that $ P\Omega  = 0, $
which represents the vacuum. The vector $ \Omega $ satisfies the
Reeh-Schlieder property,
i.e. the sets of vectors $ {\cal A}({\cal O})\Omega $ and\footnote{$
^{3)}
$}{{\sevenrm Given a set $ {\cal C} $ of
bounded operators on $ {\cal H}, $ the symbol $ {\cal C}^{\prime} $
denotes
the set of bounded operators
commuting with all elements of $ {\cal C}. $}} $ {\cal A}({\cal
O})^{\prime} \Omega $ are dense in $ {\cal H} $ for every $ {\cal O}. $

\vskip 12pt

4. {\sl (Nuclearity)\/} Let $ H = P^\mu \cdot e_\mu $ be the
Hamiltonian in a
given Lorentz system with
time-direction fixed by the positive timelike vector $ e \in  V_+, $ $
e^2=1\
; $ let $ \beta  > 0 $
and let $ \theta_{ \beta ,{\cal O}}: $ $ {\cal A}({\cal O})
\longrightarrow
{\cal H} $ be the linear mapping
$$ \theta_{ \beta ,{\cal O}}(A) = {\rm e}^{-\beta H}A\ \Omega,\ \ A \in
{\cal
 A}({\cal O})\ . \eqno (2.4) $$
This mapping is nuclear for every $ {\cal O} $ and $ \beta  > 0. $ This
means
that for fixed $ \beta $
and $ {\cal O} $ there is a sequence of vectors $ \Phi_ i\in{\cal H} $
and of
bounded linear functionals $ \varphi_ i $
on $ {\cal A}({\cal O}) $ such that $ \sum^{ }_ i \left\Vert \varphi_ i
\right\Vert \left\Vert \Phi_ i \right\Vert  < \infty $ and
$$ \theta_{ \beta ,{\cal O}}(A) = \sum^{ }_ i\varphi_ i(A)\cdot
\Phi_ i,\ \ \ \ \ A \in  {\cal A}({\cal O})\ . \eqno (2.5) $$
The nuclear norm of $ \theta_{ \beta ,{\cal O}} $ is defined by
$$ \left\Vert \theta_{ \beta ,{\cal O}} \right\Vert_ 1 = {\rm inf}
\sum^{ }_ i
\left\Vert \varphi_ i \right\Vert \left\Vert \Phi_ i \right\Vert \ ,
\eqno (2.6) $$
where the infimum is to be taken with respect to all decompositions of $
\theta_{ \beta ,{\cal O}} $
of the form (2.5). One then postulates that, for sufficiently small $
\beta  >
0 $
and large balls $ {\cal O} $ with radius $ r, $ the following
bound holds:
$$ \left\Vert \theta_{ \beta ,{\cal O}} \right\Vert_ 1 \leq  {\rm
e}^{cr^m\beta^{ -n}}\ , \eqno (2.7) $$
where $ c, $ $ m, $ $ n $ are positive numbers which do not depend on $
r $
and $ \beta. $

As it was discussed in $\lbrack$BW$\rbrack$, the nuclear norm $
\left\Vert
\theta_{ \beta ,{\cal O}} \right\Vert_ 1 $ may be regarded as a
substitute to the partition function of the theory in finite volume at
temperature $ \beta^{ -1}. $ The bound (2.7) may thus be understood as
the
requirement
that the free energy of the system exhibits a regular behaviour in the
thermodynamic limit and at large temperatures. In fact, one expects
that in
physically reasonable theories one can put $ m, $ $ n $ equal to the
dimension
of
space (Stefan-Boltzmann law). It is note-worthy that the nuclearity
condition
holds in all Lorentz systems if it holds in some.

Having listed the relevant properties of the underlying theory in terms
of
the vacuum sector, let us now recall from $\lbrack$BJ$\rbrack$ how one
one can
proceed from
this sector to thermal equilibrium states at temperature $ \beta^{
-1}>0. $
Let $ {\cal O}_r $ be the
ball of radius $ r $ centered at the origin of $ \Bbb R^d $ and let
$$ \Lambda  = \left({\cal O}_r,{\cal O}_R \right),\ \ \ \ r < R
\eqno (2.8) $$
be any pair of such balls. (In order not to overburden the notation
we do not indicate the dependence of $ \Lambda $ on $ r $ and $ R). $ It
follows from the above
assumptions that, for every $ \Lambda , $ there exists a product vector
$ \eta_ \Lambda \in{\cal H} $ with the
factorization property
$$ \left(\eta_ \Lambda ,AB^{\prime} \eta_ \Lambda \right) =
(\Omega ,A\Omega)
\left(\Omega ,B^{\prime} \Omega \right)\ \ {\rm for} \ \ \ A\in{\cal A}
\left({\cal O}_r \right)\ ,\ \ B^{\prime} \in{\cal A}
\left({\cal O}_R \right)^{\prime} \ . \eqno (2.9) $$
The vector $ \eta_\Lambda $ is not completely fixed by this equation
and it can be
modified by applying any isometry in $ {\cal A} \left({\cal O}_r
\right)^{\prime}  \wedge  {\cal A}
\left({\cal O}_R \right)^{\prime\prime} . $
Yet there is a canonical
choice for it in the so-called natural cone $ P^{\natural} $ affiliated
with $
\Omega $ and with the
von\nobreak\ Neumann algebra $ {\cal A} \left({\cal O}_r
\right)^{\prime\prime} \vee{\cal A} \left({\cal O}_R \right)^{\prime} .
$ The
latter fact was used in some arguments
in $\lbrack$BJ$\rbrack$, but it is of no relevance in the sequel. From
the
physical
viewpoint, the vector $ \eta_ \Lambda $ describes a state in which no
correlations exist
between the finite region $ {\cal O}_r $ and the causal complement of $
{\cal
O}_R. $ The ambiguities
on its determination correspond to the variety of possible boundary
conditions for systems in a finite volume, but the general arguments
developed below are in fact independent of this variety of situations.
It is
only in the passage to the thermodynamic limit (done in Sec.\ 4) that we
shall
be led to exhibit (in a certain sense) the role of the boundary
conditions in
the derivation of the final result.

With the help of the vectors $ \eta_ \Lambda $ one constructs subspaces
$
{\cal H}(\Lambda)  \doteq  \charlvupup{ \dasharrowfill}{{\cal A}
\left({\cal
O}_r \right)\eta_ \Lambda}  \subset  {\cal H}, $
where the bar denotes closure. The vectors in $ {\cal H}(\Lambda) $
describe
excitations of
the vacuum which are strictly localized in $ {\cal O}_R $ and exhaust
all
partial states
on $ {\cal A} \left({\cal O}_r \right). $ Thus the spaces $ {\cal
H}(\Lambda)
$ replace in the present setting the state
spaces which arise in finite volume theories. Let us denote by $
E(\Lambda) $
the orthogonal
projection onto $ {\cal H}(\Lambda) ; $ it has been shown in
$\lbrack$BJ$\rbrack$ that $ E(\Lambda) $ converges to 1 if
the radii $ r, $ $ R $ of the underlying balls approach infinity in an
appropriate manner. Moreover, there holds the following crucial lemma
$\lbrack$BJ$\rbrack$.

\vskip 12pt

\noindent \underbar{Lemma 2.1}\nobreak\ : {\it The operators $
E(\Lambda) {\rm
e}^{-\beta H} $ are of trace class for every $ \Lambda = \left({\cal
O}_r,{\cal O}_R \right), $ $ \beta  > 0 $
and there holds
$$ {\rm Tr} \left\vert E(\Lambda) {\rm e}^{-\beta H} \right\vert  \leq
{\rm
e}^{cR^m\beta^{ -n}}\ . \eqno (2.10) $$
Here $ c, $ $ m, $ $ n $ are the constants appearing in the nuclearity
condition and $ R, $ $ {\beta}^{-1} $
have to be sufficiently large.}

\vskip 12pt

This result allows one to define on $ {\cal H} $ the density matrices
$$ \rho_{ \beta ,\Lambda}  \doteq  {1 \over Z_{\beta ,\Lambda}}  \cdot
E(\Lambda
)  {\rm e}^{-\beta H}E(\Lambda) \ , \eqno (2.11) $$
where
$ Z_{\beta ,\Lambda} $ is a normalization factor entailing that ${\rm
Tr} \
\rho_{ \beta ,\Lambda}  = 1. $ The corresponding states $ \omega_{
\beta ,\Lambda} , $ given by $$ \omega_{ \beta ,\Lambda}( A) = {\rm Tr}
\ \rho_{ \beta ,\Lambda} A\ ,\ \ \ \ \ A \in  {\cal A}\ , \eqno (2.12)
$$
satisfy a local version of the KMS-condition $\lbrack$BJ$\rbrack$ which
indicates that these
states are close to thermal equilibrium in the region $ {\cal O}_r. $
Moreover, by making
use of the fact that $ E(\Lambda) $ tends to 1 in the limit of large $
r,R $
it has been
shown in $\lbrack$BJ$\rbrack$ that the states $ \omega_{ \beta
,\Lambda} $
have limit points
$$ \omega_ \beta  = \lim_{i} \omega_{ \beta ,\Lambda_ i} \eqno (2.13) $$
which satisfy the KMS-condition at temperature $ \beta^{ -1} $ with
respect to
the time
translations in the given Lorentz frame. The states $ \omega_ \beta $
thus
describe systems
at thermal equilibrium $\lbrack$HKTP$\rbrack$, $\lbrack$PW$\rbrack$.

In the subsequent section we will analyse the properties of the
approximating
states $ \omega_{ \beta ,\Lambda} $ more closely in order to gain
further
information on their limits
$ \omega_ \beta . $

\vskip 24pt

\noindent {\bf 3. Implications of the relativistic spectrum condition}

\vskip 12pt

\noindent We will now study the consequences of the relativistic
spectrum condition for
the properties of the approximating states $ \omega_{ \beta ,\Lambda} .
$ In a
first step we will show
that the correlation functions of these states satisfy a local version
of the
relativistic KMS-condition, as outlined in the Introduction.

\vskip 12pt

\noindent \underbar{Lemma 3.1.} : {\it Let $ \Lambda = \left({\cal
O}_r,{\cal
O}_R \right), $ $ \beta  >0 $ and let $ \omega_{ \beta ,\Lambda} $ be
the
corresponding
state defined in relation (2.12). For any $ \rho  < r $ and any pair of
operators $ A,B\in{\cal A} \left({\cal O}_\rho \right) $
there exists a function $ F(z) $ which is analytic in the tube $ {\cal
T}_{\beta e}= \left\{ z\in \Bbb C^d: \right. $ $ {\rm Im} \ z \in
V_+\cap
\left. \left(\beta e+V_- \right) \right\} $
and continuous at the boundaries $ {\rm Im} \ z = 0, $ $ {\rm Im} \
z=\beta e
$ with boundary values
given by
$$ F(x) = \omega_{ \beta ,\Lambda} \left(A\ \alpha_ x(B) \right)\ \ \ \
{\rm
and} \ \ \ \  F(x+i\beta e)= \omega_{ \beta ,\Lambda} \left(\alpha_
x(B)A
\right) $$
for $ \vert x\vert  < r-\rho . $}

\vskip 12pt

\noindent \underbar{Proof}\nobreak\ : The relativistic spectrum
condition
implies that the
operator functions $ z \longrightarrow  {\rm e}^{izP} $ and $ z
\longrightarrow  {\rm e}^{-izP-\beta H} $ are analytic on $ {\cal
T}_{\beta
\cdot e} $ and continuous on $ \charlvupup{ \dasharrowfill}{{\cal
T}_{\beta
\cdot e}} $ in the
strong operator topology. Moreover, since\footnote{$ ^{4)} $}{{\sevenrm
In the
following, we
introduce proper coordinates and denote the time and space components of
(complex) $ d $-vectors $ w $ by $ \left(w_0, {\bf w} \right).
$}} $ \left\vert {\rm e}^{i w P} \right\vert  \leq
{\rm e}^{- ( {\rm Im} w_0 \  - \  \vert {\rm Im} {\bf w} \vert )
H} $ it follows from
Lemma 2.1 that the function
$$ z \longrightarrow  F(z) = {1 \over Z_{\beta ,\Lambda}}  {\rm Tr} \
E(\Lambda)  {\rm e}^{izP} B\ {\rm e}^{-izP-\beta H}E(\Lambda) A $$
is well-defined on $ \charlvupup{ \dasharrowfill}{{\cal T}_{\beta \cdot
e}}. $
If we denote by $ X $ the operator under the trace, there
holds for $ z \in  \charlvupup{ \dasharrowfill}{{\cal T}_{\alpha \cdot
e}}, $
$ 0 < \alpha  < \beta , $ the following operator inequality:
$$ \left(X^\ast X \right)^{1/2} \leq  \Vert B\Vert  \cdot  \left\vert
{\rm
e}^{-(\beta -\alpha) H}E(\Lambda) A \right\vert \ , $$
and similarly for $ z \in  \charlvupup{ \dasharrowfill}{{\cal
T}_{\alpha \cdot
e}} + i(\beta -\alpha) e, $ $ 0<\alpha <\beta , $
$$ \left(XX^\ast \right)^{1/2} \leq  \Vert A\Vert\Vert B\Vert \cdot
\left\vert
{\rm e}^{-(\beta -\alpha) H}E(\Lambda) \right\vert \ . $$
Thus, because of the uniformity of these bounds in $ z $ and of Lemma
2.1 one
may
conclude $\lbrack$K, Chap.7$\rbrack$ that the function $ F $ is
analytic in $
{\cal T}_{\beta e} $ and continuous
at all boundary points $ {\rm Im} \ z = 0 $ and $ {\rm Im} \ z = i\beta
e. $

The calculation of the boundary values of $ F $ is accomplished by
making use
of
the cyclicity of the trace and of the fact that the projection $
E(\Lambda) $
commutes
with all operators in $ {\cal A} \left({\cal O}_r \right). $ In fact,
for $
{ B }\in  {\cal A} \left({\cal O}_\rho \right) $ and $ \vert x\vert
 <
r-\rho $ there
holds
$$ \eqalign{ F(x) & = {1 \over Z_{\beta ,\Lambda}}  {\rm Tr} \
E(\Lambda)
\alpha_ x(B) {\rm e}^{-\beta H}E(\Lambda) A \cr  &  = {1 \over Z_{\beta
,\Lambda}}  {\rm Tr} \ E(\Lambda)  {\rm e}^{-\beta H}E(\Lambda)  A\
\alpha_
x(B) = \omega_{ \beta ,\Lambda} \left(A\ \alpha_ x(B) \right)\ , \cr} $$
and similarly
$$ \eqalign{ F(x+i\beta e) & = {1 \over Z_{\beta ,\Lambda}}  {\rm Tr} \
E(\Lambda)  {\rm e}^{-\beta H} \alpha_ x(B) E(\Lambda) A \cr  &  = {1
\over
Z_{\beta ,\Lambda}}  {\rm Tr} \ E(\Lambda)  {\rm e}^{-\beta H}E(\Lambda)
\alpha_ x(B)A = \omega_{ \beta ,\Lambda} \left(\alpha_ x(B)A \right)\ .
\cr}
$$
This completes the proof of the statement.

\vskip 10pt

This result is a first indication that the thermal equilibrium states $
\omega_ \beta $
arising as limit points of the family of states $ \omega_{ \beta
,\Lambda} $
for large $ r,R $ satisfy
the relativistic KMS-condition, since the restrictions on the
localization
properties of the operators $ A,B $ and the size of $ \vert x\vert $
disappear
in this limit.
We will discuss in the next section conditions on the approximating
states $
\omega_{ \beta ,\Lambda} $
which allow one to establish this fact rigorously.

Before we enter into that discussion, let us complement the preceding
lemma by
a result of a similar nature. We shall make use of the fact that any
positive
linear functional on $ {\cal A} $ induces, by the GNS-construction, a
representation of
$ {\cal A} $ on some Hilbert space. The functionals of interest here
are $
\omega_{ \beta ,\Lambda} ; $ so there
exists a Hilbert space $ {\cal H}_{\beta ,\Lambda} , $ a distinguished
unit
vector $ \Omega_{ \beta ,\Lambda} \in{\cal H}_{\beta ,\Lambda} , $ and a
homomorphism $ \pi_{ \beta ,\Lambda} $ of $ {\cal A} $ into the algebra
of
bounded operators on $ {\cal H}_{\beta ,\Lambda} $ such
that
$$ \omega_{ \beta ,\Lambda}( A) = \left(\Omega_{ \beta ,\Lambda} ,\pi_{
\beta
,\Lambda}( A)\Omega_{ \beta ,\Lambda} \right)\ \ \ \ {\rm for} \ \ \ A
\in
{\cal A}\ . \eqno (3.1) $$
In the subsequent lemma we make use of a more concrete realization of
this
representation.

\vskip 24pt

\noindent \underbar{Lemma 3.2} : {\it Let $ \Lambda  = \left({\cal
O}_r,{\cal
O}_R \right), $ $ \beta  > 0 $ and $ \rho  < r. $ Then the vector
valued functions
$$ x \longrightarrow  \pi_{ \beta ,\Lambda} \left(\alpha_ x(A)
\right)\Omega_{
\beta ,\Lambda},\ \ \ \ \ A \in  {\cal A} \left({\cal O}_\rho \right) $$
can be continued analytically from the real ball $ \vert x\vert  <
r-\rho $
into the domain $ {\cal T}_{{\beta \over 2}e}, $
and they are weakly continuous at the boundary sets $ {\rm Im} \ z = 0,
$ $
\vert {\rm Re} \ z\vert  < r - \rho , $
and $ {\rm Im} \ z = \beta /2\cdot e, $ $ {\rm Re} \ z \in  \Bbb R^d. $}

\vskip 10pt

\noindent \underbar{Remark} : These functions are also strongly
continuous at
the
boundary points if $ x \longrightarrow  \alpha_ x(A) $ is
differentiable in
norm.

\vskip 12pt

\noindent \underbar{Proof} : We begin by recalling the standard
representation
of $ {\cal A} $
induced by density matrices in the vacuum sector $ {\cal H}. $ Let $
{\cal K}
$ be the space of
Hilbert-Schmidt operators on $ {\cal H}, $ equipped with the scalar
product
$$ \left\langle K_1\mid K_2 \right\rangle  = {\rm Tr} \ K^\ast_ 1K_2\ \
\ {\rm
for} \ \ \ K_1,K_2 \in  {\cal K}\ , $$
and let $ \pi $ be the representation of $ {\cal A} $ which acts on $
{\cal K}
$ by left multiplication,
$$ \pi( A) K = AK\ \ \ {\rm for} \ \ \ A \in  {\cal A},\ \ K \in  {\cal
K}\ .
$$
Then there holds
$$ \left\langle K_1\mid \pi( A)K_2 \right\rangle  = {\rm Tr} \ K^\ast_
1A\ K_2
= {\rm Tr} \ K_2K^\ast_ 1A\ . $$
Within this setting one can identify the vector $ \Omega_{ \beta
,\Lambda} $
with the trace class
(hence Hilbert-Schmidt class) operator $ Z^{-1/2}_{\beta ,\Lambda}
E(\Lambda)
{\rm e}^{-\beta /2\ H}, $ the Hilbert space $ {\cal H}_{\beta ,\Lambda}
$
with the subspace $ \charlvupup{ \dasharrowfill}{{\cal A}\ E(\Lambda)
{\rm
e}^{-\beta /2\ H}} $ of $ {\cal K}, $ and the representation $ \pi_{
\beta
,\Lambda} $ with the
restriction of $ \pi $ to this subspace. So, for the proof of the
statement,
we have
to study the properties of the function (taking its values in $ {\cal
K}) $
$$ x \longrightarrow  \alpha_ x(A) E(\Lambda)  {\rm e}^{-\beta /2\ H}
,\ \
\ \ \ A \in  {\cal A} \left({\cal O}_\rho \right) $$
which coincides for $ \vert x\vert  < r-\rho $ with
$$ x \longrightarrow  E(\Lambda)  \alpha_ x(A) {\rm e}^{-\beta /2\ H}\
. $$
The latter function can be extended (as a function with values in $
{\cal K})
$ to the domain $ {\cal T}_{\beta /2\cdot e}, $
by setting
$$ z \longrightarrow K(z) = E(\Lambda)  {\rm e}^{izP}A\ {\rm
e}^{-izP-\beta
/2\ H}\ . $$
In view of the arguments in the preceding lemma, it is apparent that $
K(z) $
is
analytic on $ {\cal T}_{{\beta \over 2}e} $ and continuous at the
boundary set
$ {\rm Im} \ z = i\beta /2\cdot e $ of this
domain. It requires more work to prove that it is also continuous at
the real
boundary points $ {\rm Im} \ z = 0, $ $ \vert {\rm Re} \ z\vert  <
r-\rho . $

The spectrum condition implies that $ z \longrightarrow K(z), $
regarded as an
element of the
space of bounded operator-valued functions, is continuous at the real
boundary points in the strong operator topology. Hence if $ K_0\in{\cal
K} $
is any
operator of {\sl finite rank\/}, the function
$$ z \longrightarrow  \left\langle K_0\mid K(z) \right\rangle  = {\rm
Tr} \
K^\ast_ 0E(\Lambda) {\rm e}^{izP}A\ {\rm e}^{-izP-\beta /2\ H} $$
is continuous at the real boundary set. But the finite rank operators
are
dense
in $ {\cal K}; $ so the proof is complete if we can show that $ K(z) $
is
uniformly bounded
in $ {\cal K} $ in a neighbourhood of the boundary points. To verify
this, we
pick any
real $ x, $ $ \vert x\vert  < r-\rho , $ and any $ y \in  V_+ $ with $
y_0 +
\vert \underline{y}\vert  = \varepsilon( x) = {\rm min}(\beta /2,r-\rho
-\vert x\vert) . $
Keeping these data fixed for a moment, we consider for $ w \in  \Bbb C
$ with
$ 0 \leq  {\rm Re} \ w \leq  1, $
$ \vert {\rm Im} \ w\vert  \leq  1 $ the function
$$ w \longrightarrow  K_{x,y}(w) = K(x+iwy)\ . $$
This function is analytic as a function with values in $ {\cal K} $ in
the
interior of its
definition domain, as a
consequence of the analyticity properties of $ z \longrightarrow  K(z).
$ By
making use of the
restrictions on $ x,y, $ of the spectrum condition and of Lemma 2.1, we
get
for $ 0 < {\rm Re} \ w \leq  1 $
the estimate
$$ \eqalign{ {\rm Tr} \left\vert K_{x,y}(w) \right\vert &  \leq  \Vert
A\Vert
\cdot  {\rm Tr} \left\vert E(\Lambda) {\rm e}^{- {\rm Re} \ w
\left(y_0-\vert
\underline{y}\vert \right)H} \right\vert \cr  &  \leq  \Vert A\Vert
\cdot
{\rm e}^{cR^m \left( {\rm Re} \ w \left(y_0-\vert \underline{y}\vert
\right)
\right)^{-n}}\ . \cr} $$
On the other hand, if $ {\rm Re} \ w = 0, $ $ \vert {\rm Im} \ w\vert
\leq 1 $
we get, by using again the
fact that $ E(\Lambda) $ commutes with the elements of $ {\cal A}
\left({\cal
O}_r \right), $
$$ {\rm Tr} \left\vert K_{x,y}(w) \right\vert  \leq  \Vert A\Vert
\cdot  {\rm
Tr} \left\vert E(\Lambda) {\rm e}^{-\beta /2\ H} \right\vert \ . $$
These crude a priori bounds suffice to complete the proof by an
argument of
Phragm\'en-Lindel\"of type. Consider the region
$$ D = \left\{ w \in  \Bbb C\ : 0 \leq  {\rm Re} \ w \leq  (1-\vert
{\rm Im} \
w\vert)^ 2 \right\} $$
and the auxiliary function on $ D $ given by
$$ a(w) = {\rm exp} \left(- {\rm e}^{1/(1+iw)}- {\rm e}^{1/(1-iw)}+2e
\right)\
. $$
This function is analytic on the interior of $ D, $ continuous at the
boundary,
and $ a(\pm i)=0. $ Moreover, there holds for real $ w $
$$ a(w) \geq  1,\ \ \ \ \ 0 \leq  w \leq  1\ . $$
We multiply $ K_{x,y} $ with this function in order to suppress a
possibly
singular
behaviour of $ K_{x,y} $ in the neighbourhood of the imaginary axis. By
using
the
preceding bounds on $ {\rm Tr} \left\vert K_{x,y}(w) \right\vert , $ we
obtain
for all $ x,y $ as specified above the
following estimate on the boundary of $ D $
$$ \sup_{w\in \partial D}\vert a(w)\vert {\rm Tr} \left\vert K_{x,y}(w)
\right\vert  \leq  M \left(y_0-\vert \underline{y}\vert \right)\ , $$
where $ u \longrightarrow M(u), $ $ u>0 $ is continuous and
monotonically
decreasing. We pick now
any $ K_0\in{\cal K} $ of finite rank and consider the function
$$ w \longrightarrow  \left\langle K_0 \left\vert a(w)K_{x,y}(w) \right.
\right\rangle $$
which is continuous in $ D $ and analytic in the interior. By setting $
\Vert
K\Vert_ 2=\langle K\mid K\rangle^{ 1/2}, $
$ K\in{\cal K}, $ and taking into account that $ \Vert K\Vert_ 2 \leq
{\rm
Tr}\vert K\vert , $ it follows from the maximum
modulus principle and from the preceding estimate that, for $ w\in D $
$$ \left\vert \left\langle K_0 \left\vert a(w)K_{x,y}(w) \right.
\right\rangle
\right\vert  \leq  M \left(y_0-\vert \underline{y}\vert \right)\cdot
\left\Vert K_0 \right\Vert_ 2\ . $$

Since the operators $ K_0 $ of finite rank are dense in $ {\cal K}, $
this
implies in
particular that for real $ w, $ $ 0 \leq  w \leq  1, $
$$ \Vert K(x+iwy)\Vert_ 2 = \left\Vert K_{x,y}(w) \right\Vert_ 2 \leq
{1
\over a(w)} M \left(y_0-\vert \underline{y}\vert \right) \leq  M
\left(y_0-\vert \underline{y}\vert \right)\ . $$
By bearing in mind the special choice of $ y, $ we conclude from this
estimate
that
$$ \Vert K(x+iy)\Vert_ 2 \leq  M \left(\varepsilon( x) {y_0-\vert
\underline{y}\vert \over y_0+\vert \underline{y}\vert} \right) $$
for all $ x, $ $ \vert x\vert  < r-\rho , $ and $ y\in V_+, $ $
y_0+\vert
\underline{y}\vert  \leq  \varepsilon( x) = {\rm min} \left(
\beta / 2, \  r-\rho -\vert x\vert \right). $ This
completes the
proof of the weak continuity of $ z \longrightarrow  K(z) $ at the real
boundary points.

\vskip 10pt

We conclude this section with the remark that, for all $ \rho  < r, $
the
vector $ \Omega_{ \beta ,\Lambda} $ is separating
for the algebra $ \pi_{ \beta ,\Lambda} \left({\cal A} \left({\cal
O}_\rho
\right) \right)^-. $ This can most easily be seen in the standard
representation used in the proof of the preceding lemma: if $ Z\cdot
\Omega_{
\beta ,\Lambda} =0 $ for
some $ Z\in \pi_{ \beta ,\Lambda} \left({\cal A} \left({\cal O}_\rho
\right)
\right)^-, $ then $ Z\ E(\Lambda) {\rm e}^{-\beta /2\ H}=0 $ and
consequently
$ Z\ E(\Lambda) =0 $ since $ {\rm e}^{-\beta /2\ H} $
is invertible. Moreover, $ \left[\alpha_ x(Z),E(\Lambda) \right]=0 $ if
$
\vert x\vert  < r-\rho , $ hence $ Z=0 $ by a theorem
of Schlieder $\lbrack$Sch$\rbrack$.

\vskip 24pt

\noindent {\bf 4. The role of boundary effects and the thermodynamic
limit}

\vskip 12pt

\noindent After having seen how analyticity properties of the
correlation functions
affiliated with the approximating states $ \omega_{ \beta ,\Lambda} $
result
from the relativistic
spectrum condition, we now turn to the formulation of conditions which
imply
that these properties persist in the limit states $ \omega_ \beta . $

Our starting point is the heuristic idea that the restrictions of $
\omega_
\beta $ and $ \omega_{ \beta ,\Lambda} $
to any given local algebra $ {\cal A}({\cal O}) $ should practically
look
alike if the regions $ {\cal O}_r, $
$ {\cal O}_R $ are sufficiently large compared to $ {\cal O}. $ Any
differences between the
restricted (partial) states should be due to boundary effects in the
state $
\omega_{ \beta ,\Lambda} $
which are localized in a layer $ \partial \Lambda $
in the spacelike (causal) complement of ${\cal O}_r, $ where the
equilibrium situation
in $ {\cal O}_r $ is decoupled from the exterior vacuum (cf. the
definition of
$ {\cal H}(\Lambda) $ and
relation (2.9)).

One may expect that these boundary effects are removable in generic
cases\footnote{$ ^{5)} $}{{\sevenrm Boundary effects play a prominent
role at
phase transition
points, where our arguments are less conclusive.}} by operations in the
layer $ \partial \Lambda $ and hence {\sl a fortiori\/} by operations
in the
causal complement of $ {\cal O}. $
The latter operations can be described in a relativistic theory by
operators
$ T $ which commute with all observables in $ {\cal O}. $

To substantiate this idea, let us assume that $ \omega_ \beta
\upharpoonright
{\cal A}({\cal O}) $
(namely, the
restriction of $ \omega_ \beta $ to $ {\cal A}({\cal O})) $ is normal
with
respect to $ \omega_{ \beta ,\Lambda}  \upharpoonright {\cal A}({\cal
O}). $
By taking into account the remark at the end of the preceding section,
it
follows
$\lbrack$D$\rbrack$ that $ \omega_ \beta  \upharpoonright {\cal
A}({\cal O}) $
is a vector state in the GNS-representation
$ (\pi_{ \beta ,\Lambda} , $ $ {\cal H}_{\beta ,\Lambda} , $ $ \Omega_{
\beta
,\Lambda} ) $ induced by $ \omega_{ \beta ,\Lambda}  \upharpoonright
{\cal
A}({\cal O}), $ i.e. there is a
vector $ \Omega_ \beta \in{\cal H}_{\beta ,\Lambda} $ such that $
\omega_
\beta( A) = \left(\Omega_ \beta ,\pi_{ \beta ,\Lambda}( A)\Omega_ \beta
\right) $ for $ A\in{\cal A}({\cal O}). $ It is therefore
possible $\lbrack$S, Chap.2.7$\rbrack$ to introduce a linear operator $
T $ in
$ {\cal H}_{\beta ,\Lambda} , $ by setting
$$ T\cdot \pi_{ \beta ,\Lambda}( A)\Omega_{ \beta ,\Lambda}  = \pi_{
\beta
,\Lambda}( A)\Omega_ \beta \ \ \ \ \ \ {\rm for} \ \ \ \ \ \  A \in
{\cal
A}({\cal O}). \eqno (4.1) $$
This operator is well defined on the dense domain $ \pi_{ \beta
,\Lambda}({\cal A}({\cal O}))\Omega_{ \beta ,\Lambda} , $ as it follows
likewise from the remark at the end of the preceding section, and it is
also
apparent from (4.1) that this operator commutes on its domain with the
elements of the
algebra $ \pi_{ \beta ,\Lambda}({\cal A}({\cal O})). $ Hence $ T $ has
the
heuristically expected properties of an
operator which removes boundary effects.

We propose to classify the strength of boundary effects by the
continuity
properties of $ T. $ Depending on the nature of these effects, $ T $
may be a
bounded
operator, a closable unbounded operator or even, in some instances, a
non-closable operator. In the following criterion we distinguish two
generic
cases.

\vskip 12pt

\noindent \underbar{Criterion} : {\it The state $ \omega_ \beta $ is
said to
be {\sl strongly resistant\/} to
boundary effects if for each bounded region $ {\cal O} $ there is a $
\Lambda
= \left({\cal O}_r,{\cal O}_R \right) $ and a
bounded operator $ T \in  \pi_{ \beta ,\Lambda}({\cal A}({\cal
O}))^{\prime} $
such that
$$ \omega_ \beta( A) = \left(T\Omega_{ \beta ,\Lambda} ,\pi_{ \beta
,\Lambda}(
A)T\Omega_{ \beta ,\Lambda} \right)\ \ \ \ {\rm for} \ \ \ \ A \in
{\cal
A}({\cal O})\ . $$
It is said to be {\sl resistant\/} if $ T $ is a closable unbounded
operator
which
commutes on its domain $ \pi_{ \beta ,\Lambda}({\cal A}({\cal
O}))\Omega_{
\beta ,\Lambda} $ with all the elements of $ \pi_{ \beta ,\Lambda}({\cal
A}({\cal O})) $ and
satisfies $ \left\Vert T^\ast T\Omega_{ \beta ,\Lambda} \right\Vert  <
\infty
. $}

It is of interest here that these conditions can be reformulated in
terms of
the underlying functionals $ \omega_{ \beta ,\Lambda} $ and $ \omega_
\beta ,
$ thereby permitting a different
physical interpretation.

\vskip 12pt

\noindent \underbar{Lemma 4.1} : {\it The state $ \omega_ \beta $ is
strongly
resistant to boundary effects
iff for each bounded region $ {\cal O} $ there is a $ \Lambda  =
\left({\cal
O}_r,{\cal O}_R \right) $ and a positive constant
$ c $ such that
$$ \omega_ \beta \left(A^\ast A \right) \leq  c\cdot \omega_{ \beta
,\Lambda}
\left(A^\ast A \right) \ ,\ \ \ \ \ \ A \in  {\cal A}({\cal O}) \ .
\eqno (4.2) $$
It is resistant iff
$$ \left\vert \omega_ \beta( A) \right\vert^ 2 \leq  c\cdot \omega_{
\beta
,\Lambda} \left(A^\ast A \right) ,\ \ \ \ \ A \in  {\cal A}({\cal
 O})\
. \eqno (4.3) $$
}

\vskip 12pt

\noindent \underbar{Proof} : We begin by proving the second part of the
statement: it is
a direct consequence of condition (4.3) that the state $ \omega_ \beta
\upharpoonright {\cal A}({\cal O}) $
is normal with respect to $ \omega_{ \beta ,\Lambda}  \upharpoonright
{\cal
A}({\cal O}). $ Hence, as explained
before, there is a vector $ \Omega_ \beta  \in  {\cal H}_{\beta
,\Lambda} $
inducing the state $ \omega_ \beta  \upharpoonright {\cal A}({\cal O}) $
in the representation $ \left(\pi_{ \beta ,\Lambda} , {\cal H}_{\beta
,\Lambda} , \Omega_{ \beta ,\Lambda} \right), $ and a linear operator $
T $
such that
relation (4.1) holds. It also follows from (4.3) that for any $ A,B \in
 {\cal
A}({\cal O}) $
$$ \matrix{ \left\vert \left(\pi_{ \beta ,\Lambda}( B)\Omega_ \beta
,\pi_{
\beta ,\Lambda}( A)\Omega_ \beta \right) \right\vert^ 2 = \left\vert
\omega_
\beta \left(B^\ast A \right) \right\vert^ 2 \leq  c\ \omega_{ \beta
,\Lambda}
\left(A^\ast BB^\ast A \right) \cr \leq  c \left\Vert BB^\ast
\right\Vert
\omega_{ \beta ,\Lambda} \left(A^\ast A \right) = c\Vert B\Vert^ 2
\left\Vert
\pi_{ \beta ,\Lambda}( A)\Omega_{ \beta ,\Lambda} \right\Vert^ 2\ ,
\cr} $$
and consequently
$$ \left\vert \left(\pi_{ \beta ,\Lambda}( B)\Omega_ \beta ,T\ \pi_{
\beta
,\Lambda}( A)\Omega_{ \beta ,\Lambda} \right) \right\vert  \leq
c^{1/2}\Vert
B\Vert \left\Vert \pi_{ \beta ,\Lambda}( A)\Omega_{ \beta ,\Lambda}
\right\Vert \ . $$
This inequality shows that the adjoint $ T^\ast $ is defined on the
range of $
T, $ hence
$ T $ is closable. By setting $ B=1 $ in the latter inequality, it is
also
clear that
$ \left\Vert T^\ast T\ \Omega_{ \beta ,\Lambda} \right\Vert  =
\left\Vert
T^\ast \Omega_ \beta \right\Vert  \leq  c^{1/2}, $ which proves that $
T $ has
all the properties required in our
criterion for resistance. Conversely, if $ T $ is such an operator,
there
holds
$$ \matrix{ \left\vert \omega_ \beta( A) \right\vert^ 2 =
\left\vert \left(T\Omega_{ \beta ,\Lambda} ,T\ \pi_{ \beta
,\Lambda}( A)\Omega_{ \beta ,\Lambda} \right) \right\vert^ 2 \cr \leq
\left\Vert T^\ast T\ \Omega_{ \beta ,\Lambda} \right\Vert^ 2 \left\Vert
\pi_{
\beta ,\Lambda}( A)\Omega_{ \beta ,\Lambda} \right\Vert^ 2 = \left\Vert
T^\ast
T\ \Omega_{ \beta ,\Lambda} \right\Vert^ 2\omega_{ \beta ,\Lambda}
\left(A^\ast A \right)\ , \cr} $$
which completes our proof of the second part of the lemma. The first
part
concerning strong resistance is an immediate consequence of this result.

\vskip 10pt

The first condition in the preceding lemma says that $ \omega_ \beta
\upharpoonright {\cal A}({\cal O}) $
describes a subensemble of $ \omega_{ \beta ,\Lambda}  \upharpoonright
{\cal
A}({\cal O}). $
Note that the latter state is faithful and, as such, would dominate any
other
state in the sense of relation (4.2) in locally finite theories (spin
systems). Since the nuclearity condition characterizes theories which
are in
some specific sense close to being locally finite, we believe that
condition
(4.2) is also meaningful in the present field-theoretical setting. The
second
less stringent condition amounts to the requirement that the difference
between the expectation values of any observable $ A \in  {\cal
A}({\cal O}) $
in the states $ \omega_ \beta $
and $ \omega_{ \beta ,\Lambda} , $ respectively, is dominated by the
fluctuations of $ A $ in the state $ \omega_{ \beta ,\Lambda} . $
By bearing in mind the physical situation described by $ \omega_ \beta
$ and $
\omega_{ \beta ,\Lambda} , $ this seems
to be a rather mild requirement.

Let us now turn to the discussion of the consequences of our criterion
for
the analyticity properties of the correlation functions induced by $
\omega_
\beta . $ These
functions are, according to relations (4.2) and (4.3), respectively,
and by
Lemma 3.1, dominated on the {\sl reals\/} by boundary values of analytic
functions.
Because of the underlying Hilbert space structure, it is possible to
carry
over the analyticity properties of these upper bounds to the correlation
functions.

\vskip 12pt

\noindent \underbar{Proposition 4.2} : {\it Let $ \omega_ \beta $ be a
KMS-state on $ {\cal A} $ and let $ A,B \in  {\cal A} $ be
local operators with corresponding correlation function
$$ (x_1,x_2) \longrightarrow  \omega_ \beta \left(\alpha_{
x_1}(A)\alpha_{
x_2}( B ) \right)\ . $$
If $ \omega_ \beta $ is {\sl strongly resistant\/} to boundary effects,
this
correlation function has
an analytic continuation into the domain $ -{\cal T}_{{\beta \over
2}e}\times{\cal T}_{{\beta \over 2}e}, $ and its boundary
values on $ \Bbb R^d \times  \Bbb R^d $ and $ \left(\Bbb R^d-i{\beta
\over 2}e
\right)\times \left(\Bbb R^d+i{\beta \over 2}e \right) $ are continuous.
If $ \omega_ \beta $ is (merely) {\sl resistant\/} to boundary effects,
the
correlation function has
an analytic continuation with respect to the variable $ x_1 $
(resp. $ x_2) $ into the domain $ -{\cal T}_{{\beta \over 2}e}\times
\Bbb R^d
$ (resp. $ \Bbb R^d\times{\cal T}_{{\beta \over 2}e}) $ with continuous
boundary values on $ \Bbb R^d\times  \Bbb R^d $ and $ \left(\Bbb R^d
-i{\beta
\over 2}e \right)\times  \Bbb R^d $ (resp. $ \Bbb R^d\times  \Bbb R^d $
and $ \Bbb R^d \times  \left(\Bbb R^d+i{\beta \over 2}e \right)). $}

\vskip 12pt

\noindent \underbar{Proof} : For fixed local operators $ A,B \in  {\cal
A} $
and translations $ x_1,x_2 $
varying within any given bounded subset $ {\cal R }\subset  \Bbb R^d, $
there
exists a bounded
region $ {\cal O }\subset  \Bbb R^d $ such that $ \alpha_{
x_1}(A),\alpha_{
x_2}(B) \in  {\cal A}({\cal O}). $ By assumption, there is
then a $ \Lambda  = \left({\cal O}_r,{\cal O}_R \right) $ and a
corresponding
operator $ T $ such that
$$ \matrix{ \omega_ \beta \left(\alpha_{ x_1}(A)\alpha_{ x_2}(B)
\right) =
\left(T\ \Omega_{ \beta ,\Lambda} ,\pi_{ \beta ,\Lambda} \left(\alpha_{
x_1}(A)\alpha_{ x_2}(B) \right) T\ \Omega_{ \beta ,\Lambda} \right) \cr
=
\left(T\ \pi_{ \beta ,\Lambda} \left(\alpha_{ x_1} \left(A^\ast \right)
\right)\Omega_{ \beta ,\Lambda} ,\ T\ \pi_{ \beta ,\Lambda}
\left(\alpha_{
x_2}(B) \right)\Omega_{ \beta ,\Lambda} \right)\ , \cr} $$
where we made use of the fact that $ T $ commutes with the elements of
$ \pi_{
\beta ,\Lambda}({\cal A}({\cal O})). $
If $ \omega_ \beta $ is strongly resistant to boundary effects, i.e. if
$ T $
is bounded, the
stated analyticity and continuity properties of the correlation function
follow from Lemma 3.2. (We note in this context that the {\sl joint\/}
continuity at
the boundaries in all variables is a consequence of the continuity of
the
correlation functions and of the maximum modulus principle, cf.
Appendix A.)
If
$ \omega_ \beta $  is only resistant to boundary effects, the preceding
representation of the
correlation function is not adequate since the analytic continuation of
the
underlying vector-valued functions might not remain in the domain of the
unbounded operator $ T. $ But we have
$$ \matrix{ \omega_ \beta \left(\alpha_{ x_1}(A)\alpha_{ x_2}(B)
\right) =
\left(T\ \Omega_{ \beta ,\Lambda} ,\ T\ \pi_{ \beta ,\Lambda}
\left(\alpha_{
x_1}(A)\alpha_{ x_2}(B) \right)\Omega_{ \beta ,\Lambda} \right) \cr =
\left(\pi_{ \beta ,\Lambda} \left(\alpha_{ x_1} \left(A^\ast \right)
\right)\cdot T^\ast T\ \Omega_{ \beta ,\Lambda} ,\ \pi_{ \beta ,\Lambda}
\left(\alpha_{ x_2}(B) \right)\Omega_{ \beta ,\Lambda} \right)\ , \cr}
$$
and similarly
$$ \omega_ \beta \left(\alpha_{ x_1}(A)\alpha_{ x_2}(B) \right) =
\left(\pi_{
\beta ,\Lambda} \left(\alpha_{ x_1} \left(A^\ast \right)
\right)\Omega_{ \beta
,\Lambda} ,\ \pi_{ \beta ,\Lambda} \left(\alpha_{ x_2}(B) \right)T^\ast
T\
\Omega_{ \beta ,\Lambda} \right)\ . $$
Lemma 3.2, when applied to the appropriate expression, now yields the
stated
analyticity and continuity properties in $ x_1, $ resp. $ x_2. $

\vskip 10pt

It is apparent from this argument that the conclusions hold under much
weaker
conditions. For example, the operator $ T $ in the preceding proof
could in
principle depend on the choice of the operators $ A,B \in  {\cal A}. $
One may
therefore
expect that the correlation functions exhibit analyticity properties of
the
type established in this proposition also under more general
circumstances.

\vskip 24pt

\noindent {\bf 5. Analyticity domains of correlation functions and the
relativistic KMS-condition}

\vskip 12pt

\noindent In the preceding section, we have exhibited certain specific
analyticity properties of thermal correlation functions in relativistic
quantum field
theory. By using this initial information, we will now determine the
full
primitive domains of analyticity of these functions by applying
geometrical
techniques of analytic completion, as described e.g. in
$\lbrack$BEGS$\rbrack$. Since these
techniques may be not so well-known, we have added an appendix where the
relevant notions and results are summarized.

The most general (or weakest) formulation of the expected regularity
properties of correlation functions affiliated with thermal equilibrium
states $ \omega_ \beta $ in relativistic quantum field theory is
encoded in
the following
analyticity properties.

\vskip 10pt

a) {\sl (KMS-condition)  \/ }
Let $ e \in  V_+, $ $ e^2=1 $ be the time direction of the privileged
Lorentz frame fixed
by $ \omega_ \beta $ and let $ A,B \in  {\cal A} $ be any pair of local
operators. Then there exists an
analytic function $ F_0 $ in the flat tube $ {\cal T}_{{\cal B}_0}
\subset
\Bbb C^d \times  \Bbb C^d $ with
basis\footnote{$ ^{6)} $}{{\sevenrm In the following we reserve the
letters $
x,y $ for the real
and imaginary parts of $ z \in  \Bbb C^d, $ respectively.}}
$$ {\cal B}_0 = \left\{ \left(y_1,y_2 \right) : y_1 = t_1e,\ y_2=t_2e,\
0<
t_2-t_1 < \beta \right\} \ , \eqno (5.1) $$
which is continuous on the closure of $ {\cal T}_{{\cal B}_0} $ and
has, for $
x_1,x_2\in \Bbb R^d, $
the boundary values
$$ \eqalignno{ F_0 \left(x_1,x_2 \right) & = \omega_ \beta
\left(\alpha_{
x_1}(A)\alpha_{ x_2}(B) \right) &  \cr F_0 \left(x_1-i{\beta \over 2}
e,\
x_2+i{\beta \over 2}e \right) & = \omega_ \beta \left(\alpha_{
x_2}(B)\alpha_{
x_1}(A) \right)\ . & (5.2) \cr} $$
(We note that $ F_0 $ is invariant under time-translations, $ F_0
\left(x_1+te,\ x_2+te \right)=F_0 \left(x_1,x_2 \right), $
as a consequence of this assumption, $\lbrack$BR$\rbrack$.)

\vskip 10pt

b) {\sl (Stability)  \/} For any positive timelike unit vector $ f \in
V_+, $ there exists a
function $ F_f $ which is analytic in the flat tube $ {\cal T}_{{\cal
B}_f} $
with basis
$$ {\cal B}_f = \left\{ \left(y_1,y_2 \right) : y_1=0,\ y_2=tf\ ,\ \ 0
< t <
t_f \right\} \ , \eqno (5.3) $$
$ (t_f $ being a positive number depending of $ f), $ and has continuous
boundary values
on the reals $ \Bbb R^d\times \Bbb R^d $ given by
$$ F_f \left(x_1,x_2 \right) = \omega_ \beta \left(\alpha_{
x_1}(A)\alpha_{
x_2}(B) \right)\ . \eqno (5.4) $$
(Note that by replacing $ A,B $ by $ B^\ast ,A^\ast $ and taking complex
conjugates one
obtains another function $ F^{\dagger}_ f $ which is analytic in the
flat tube
$$ {\cal T}_{{\cal B}^{\dagger}_ f} = \left\{ \left(z_1,z_2 \right) :
y_1=-tf,\ 0 < t < t_f,\ y_2=0 \right\} \eqno (5.5) $$
and coincides with $ F_f $ on the reals $ \Bbb R^d \times  \Bbb R^d.) $

Condition b) expresses the type of stability requirement presented in
the
Introduction (finiteness of local energy in all Lorentz frames). It is
of
course fulfilled when the state $ \omega_ \beta $ satisfies the
criteria of
the preceding
section (cf. Proposition 4.2). We shall now prove:

\vskip 12pt

\noindent \underbar{Proposition 5.1} : {\it Under the previous
conditions a)
and b) there
exists an analytic function $ F $ in a convex open tube $ {\cal
T}_{{\cal B}}
\subset  \Bbb C^d \times  \Bbb C^d $
with basis $ {\cal B} $ which extends the correlation function $ x_1,x_2
\longrightarrow  \omega_ \beta \left(\alpha_{ x_1}(A)\alpha_{ x_2}(B)
\right)
$
for any given local $ A,B \in  {\cal A}. $ More specifically:

\vskip 10pt

i) The basis $ {\cal B} $ is a neighbourhood in $ \Bbb R^d \times  \Bbb
R^d $
of the basis $ {\cal B}_0 $ of
condition a) and is invariant under time translations $ \left(y_1,y_2
\right)
\longrightarrow  \left(y_1+te,\ y_2+te \right), $
$ t\in \Bbb R. $ Moreover, F satisfies in its domain the invariance
condition
$$ F \left(z_1+ue,\ z_2+ue \right) = F \left(z_1,z_2 \right),\ \ \ \
u \in
 \Bbb C. $$

ii) At every real boundary point $ \left(x_1,x_2 \right) $ of $ {\cal
T}_{{\cal B}}, $ the profile $ \Lambda_{( 0,0)} $ of $ {\cal T}_{{\cal
B}} $
(corresponding to the conical boundary point $ (0,0) $ of $ {\cal B}) $
is the
cone
$$ \Lambda_{( 0,0)} = \left\{ \left(y_1,y_2 \right) : y_1 \in  V_-+te,\
y_2\in
V_++te,\ t\in \Bbb R \right\} \ , $$
and one has
$$ \lim_{\Lambda_{( 0,0)}\ni \left(\eta_ 1,\eta_ 2 \right)
\longrightarrow(
0,0)} F \left(x_1+i\eta_ 1,\ x_2+i\eta_ 2 \right) = \omega_ \beta
\left(\alpha_{ x_1}(A)\alpha_{ x_2}(B) \right)\ . $$

iii) Similarly, at every boundary point $ \left(x_1-i{\beta \over 2} e,\
x_2+i{\beta \over 2}e \right) $ of $ {\cal T}_{{\cal B}}, $ the profile
$ \Lambda_{ \left(-{\beta \over 2}e,{\beta \over 2}e \right)} $  of $
{\cal
T}_{{\cal B}} $ (corresponding to the conical boundary point $
\left(-{\beta
\over 2}e, {\beta \over 2}e \right) $ of $ {\cal B}) $
is the cone
$$ \Lambda_{ \left(-{\beta \over 2}e,{\beta \over 2}e \right)} =
\left(-{\beta
\over 2}e,{\beta \over 2}e \right) - \Lambda_{( 0,0)}\ , $$
and one has
$$ \lim_{\Lambda_{ \left(-{\beta \over 2}e,{\beta \over 2}e \right)}\ni
\left(\eta_ 1,\eta_ 2 \right) \longrightarrow \left(-{\beta \over
2}e,{\beta
\over 2}e \right)} F \left(x_1+i\eta_ 1,\ x_2+i\eta_ 2 \right) =
\omega_ \beta
\left(\alpha_{ x_2}(B)\alpha_{ x_1}(A) \right)\ . $$

In ii) and iii) the boundary values taken from the tube $ {\cal
T}_{{\cal B}}
$ exist in the
sense of continuous functions near boundary points, as specified in
Appendix
A.}

\vskip 12pt

\noindent \underbar{Proof} : The desired analytic continuation is
accomplished
in five
steps by repeated applications of the (flat) tube theorem (cf. Appendix
A). We
note that the successive continuations will be labelled by the number
of the steps where they appear.

\vskip 10pt

1. We first notice that the functions $ F_f, $ $ f \in  V_+ $ have a
common
analytic
continuation $ F_1 $ in the flat tube $ {\cal T}_{{\cal B}_1} $ whose
basis $
{\cal B}_1 $ is the convex hull of the
set
$$ \left\{ \left(y_1,y_2 \right) : y_1=0,\ y_2=tf,\ f\in V_+,\ 0 < t <
t_f
\right\} \ . $$
This follows from an iterated application of the flat tube theorem to
the
whole set of flat tubes $ {\cal T}_{{\cal B}_f}, $ with $ f $ sweeping
the set
of all directions in $ V_+, $
and a subsequent application of the tube theorem with respect to the
variable
$ z_2. $ Similarly, the functions $ F^{\dagger}_ f, $ $ f \in  V_+, $
have a
common analytic continuation
$ F^{\dagger}_ 1 $ in the flat tube $ {\cal T}_{{\cal B}^{\dagger}_ 1} $
whose basis $ {\cal B}^{\dagger}_ 1 $ is the convex hull of the set
$$ \left\{ \left(y_1,y_2 \right) : y_1=-tf,\ f\in V_+,\ 0 < t < t_f,\
y_2=0
\right\} \ . $$
At all real points, $ {\cal T}_{{\cal B}_1} $ and $ {\cal T}_{{\cal
B}^{\dagger}_ 1} $ have the respective profiles
$$ \eqalign{ \Lambda_ 1 & = \left\{ \left(y_1,y_2 \right) : y_1=0,\ y_2
\in
V_+ \right\} \cr \Lambda^{ \dagger}_ 1 & = \left\{ \left(y_1,y_2
\right) : y_1
\in  V_-,\ y_2=0 \right\} \ . \cr} $$

2. Since the functions $ F_1 $ and $ F^{\dagger}_ 1 $ coincide on the
reals $
\Bbb R^d \times  \Bbb R^d, $
they satisfy all the conditions of Lemma A.2 in the Appendix. As a
result
they have a common analytic continuation $ F_2 $ in the tube whose
basis is
the
convex hull of $ {\cal B}_1\cup{\cal B}^{\dagger}_ 1. $ The function $
F_2 $
is analytic in particular in the tube $ {\cal T}_{{\cal B}_2} $
with basis
$$ {\cal B}_2 = \left\{ \left(y_1,y_2 \right) : \left(2y_1,0 \right) \in
{\cal B}^{\dagger}_ 1,\ \left(0,2y_2 \right) \in  {\cal B}_1 \right\} $$
(corresponding to $ {\cal B}_\lambda $ with $ \lambda  =1/2 $ in Lemma
A.2).
It is clear that at each real
point $ \left(x_1,x_2 \right) $ the profile of $ {\cal T}_{{\cal B}_2}
$ is
the cone
$$ \Lambda_ 2 = \left\{ \left(y_1,y_2 \right) : y_1\in V_-,\ y_2 \in
V_+
\right\} \ , $$
and there holds the boundary relation
$$ \lim_{\Lambda_ 2\ni \left(\eta_ 1,\eta_ 2 \right) \longrightarrow(
0,0)}
F_2 \left(x_1+i\eta_ 1,\ x_2+i\eta_ 2 \right) = \omega_ \beta
\left(\alpha_{
x_1}(A)\alpha_{ x_2}(B) \right)\ . $$

\vskip 10pt

3. Next we consider the function $ x_1,x_2 \longrightarrow \omega_ \beta
\left(\alpha_{ x_2}(B)\alpha_{ x_1}(A) \right). $ An argument similar
to the one given
in the preceding steps establishes the existence of an analytic
continuation $ G $ of this function into the tube $ {\cal T}_{-{\cal
B}_2} $
with profile $ -\Lambda_ 2 $ at the
real boundary points $ \left(x_1,x_2 \right). $ For later convenience we
proceed from $ G $ to the
function $ F_3, $ given by
$$ F_3 \left(z_1,z_2 \right) \doteq  G \left(z_1+i {\beta \over 2} e,\
z_2 - i {\beta \over 2} e \right)\ , $$
which is analytic in the tube $ {\cal T}_{{\cal B}_3} $ with basis
$$ {\cal B}_3 = \left\{ \left(y_1,y_2 \right) : \left(y_1,y_2 \right)
\in
\left(- {\beta \over 2}e,{\beta \over 2}e \right) - {\cal B}_2 \right\}
$$
and profile $ \Lambda_ 3= \left(- {\beta \over 2}e,{\beta \over 2}e
\right)-\Lambda_ 2 $ at each boundary point $ \left(x_1-i{\beta \over
2}e,\
x_2+i{\beta \over 2}e \right). $ The
following boundary relation then holds:
$$ \lim_{\Lambda_ 3\ni \left(\eta_ 1,\eta_ 2 \right) \longrightarrow
\left(-
{\beta \over 2}e,{\beta \over 2}e \right)} F_3 \left(x_1+i\eta_ 1,\
x_2+i\eta_
2 \right) = \omega_ \beta \left(\alpha_{ x_2}(B)\alpha_{ x_1}(A)
\right)\ . $$

\vskip 10pt

4. So far we have only exploited the stability condition b). Now the
KMS-condition a) tells us (by an application of Lemma A.1) that, on the
one
hand,
the functions $ F_2 $ and $ F_0 $ coincide on the flat tube $ {\cal
T}_{{\cal
B}_2}\cap  {\cal T}_{{\cal B}_0} $ since they have
the same boundary value $ \omega_ \beta \left(\alpha_{ x_1}(A)\alpha_{
x_2}(B)
\right) $ on $ \Bbb R^d\times \Bbb R^d. $ On the other
hand, the functions $ F_3 $ and $ F_0 $ coincide on the flat tube $
{\cal
T}_{{\cal B}_3}\cap  {\cal T}_{{\cal B}_0} $ since they
have the same boundary value $ \omega_ \beta \left(\alpha_{
x_2}(B)\alpha_{
x_1}(A) \right) $ on $ \left(\Bbb R^d-i{\beta \over 2} e \right) \times
\left(\Bbb R^d+i{\beta \over 2}e \right). $
It follows that the functions $ F_0, $ $ F_2 $ and $ F_3 $ define a
unique
function $ F_4, $
analytic in the dumbbell-shaped tube $ {\cal T}_{{\cal B}_2\cup{\cal
B}_3\cup{\cal L}}, $ where $ {\cal L} $ is the segment $ \left\{
\left(y_1,y_2
\right) : y_2=-te,\ y_2=te,\ 0 < t < {\beta \over 2} \right\} $
in $ {\cal B}_0. $ As a result of Lemma A.3, $ F_4 $ can therefore be
analytically continued
in the convex open tube $ {\cal T}_{{\cal B}_4} $ whose basis $ {\cal
B}_4 $
is the convex hull of $ {\cal B}_2\cup{\cal B}_3\cup{\cal L}. $
Hence $ {\cal B}_4 $ contains in particular a full neighbourhood of $
{\cal L}
$ in $ \Bbb R^d\times \Bbb R^d. $

\vskip 10pt

5. In order to obtain the complete results stated in the proposition, it
remains
to establish the property of translation invariance of the function $
F_4 $
under
the transformations $ \left(z_1,z_2 \right) \longrightarrow
\left(z_1+te,\
z_2+te \right) $ and to derive its geometrical
consequences. Let
$$ N \left(z_1,z_2 \right) = { {\rm d} \over {\rm d} t} F_4 \left.
\left(z_1+te,\ z_2+te \right) \right\vert_{ t=0}\ . $$
This function is analytic in the same domain $ {\cal T}_{{\cal B}_4} $
as $
F_4 $ itself and, in view of
the remark in condition a), its boundary value on $ \Bbb R^d \times
\Bbb R^d
$
vanishes (in the sense of distributions). Therefore, in view of Lemma
A.1, $ N
$
is equal to 0 in $ {\cal T}_{{\cal B}_4}, $ which in turn implies that
$ F_4
\left(z_1+te,\ z_2+te \right)=F_4 \left(z_1,z_2 \right) $
in the whole domain $ {\cal T}_{{\cal B}_4} $ (since, in view of the
convexity
of $ {\cal T}_{{\cal B}_4}, $ each orbit $ (z_1+ue, $
$ z_2+ue), $ $ u \in \Bbb C $ has a connected intersection with $ {\cal
T}_{{\cal B}_4}). $

From this invariance property of $ F_4, $ it follows that $ F_4 $ can be
analytically
continued to an analytic function $ F $ in the tube $ {\cal T}_{{\cal
B}} $
with basis
$$ {\cal B }= \left\{ \left(y_1,y_2 \right) : y_1=\eta_ 1+te,\ y_2=\eta_
2+te,\ \left(\eta_ 1,\eta_ 2 \right)\in{\cal B}_4,\ t \in  \Bbb R
\right\} \ .
$$
Since $ {\cal B}_4 $ contains a neigbhourhood of $ {\cal L} $ (as shown
in
step 4), one easily
checks that the region
$ {\cal B} $ contains a neighbourhood of $ {\cal B}_0 $ in $ \Bbb
R^d\times
\Bbb R^d; $ moreover, the profile
of $ {\cal T}_{{\cal B}} $ at any boundary point of $ \Bbb R^d \times
\Bbb
R^d, $ (resp. $ \left(\Bbb R^d-i {\beta \over 2}e \right)\times
\left(\Bbb
R^d+i{\beta \over 2}e \right)), $
is the cone
$$ \Lambda_{( 0,0)} = \left\{ \left(y_1,y_2 \right) : y_1=\eta_ 1+te,\
y_2=\eta_ 2+te,\ \left(\eta_ 1,\eta_ 2 \right)\in \Lambda_ 2,\ t\in
\Bbb R
\right\} $$
(resp. $ \Lambda_{ \left(-{\beta \over 2}e,{\beta \over 2}e \right)} =
\left(-{\beta \over 2}e,{\beta \over 2}e \right) - \Lambda_{( 0,0)}). $
Finally, from the boundary relations
in steps 2 and 3 it follows that the function $ F $ has the boundary
values
stated in the proposition.

\vskip 10pt

Let us now specialize this proposition to the cases when the correlation
functions have the stronger analyticity properties established in the
previous
section under more restrictive assumptions. In the case when $ \omega_
\beta $
is resistant
to boundary effects (in the sense of our criterion) we have given a
direct
proof of the existence of analytic functions $ F_1 $ and $ F^{\dagger}_
1 $ in
respective flat
tubes $ {\cal T}_{{\cal B}_1} $ and $ {\cal T}_{{\cal B}^{\dagger}_ 1}
$ with
(convex) bases
$$ \eqalignno{{\cal B}_1 & = \left\{ \left(y_1,y_2 \right) : y_1=0,\
y_2 \in
V_+ \cap  \left({\beta \over 2} e + V_- \right) \right\} &  (5.6)
\cr{\cal
B}^{\dagger}_ 1 & = \left\{ \left(y_1,y_2 \right) : y_1 \in  V_- \cap
\left(-{\beta \over 2}e  + V_+ \right),\ y_2=0 \right\} &  (5.7) \cr} $$
which extend the correlation functions $ x_1,x_2 \longrightarrow
\omega_
\beta \left(\alpha_{ x_1}(A)\alpha_{ x_2}(B) \right). $ As a matter
of fact, the complete holomorphy envelope can be computed in this case
according to the following statement.

\vskip 12pt

\noindent \underbar{Proposition 5.2} : {\it If the state $ \omega_
\beta $ is
resistant to boundary effects,
then the functions $ F $ of Proposition 5.1 which extend the correlation
functions, are analytic in the (flatly bordered) convex tube $ {\cal
T}_{{\cal
B}} $ whose basis $ {\cal B} $ is defined
as follows:
$$ {\cal B}= \left\{ \left(y_1,y_2 \right) : y_1=\eta_ 1+te,\ y_2 =
\eta_
2+te,\ \left(\eta_ 1,\eta_ 2 \right) \in  {\cal B}_U,\ t \in  \Bbb R
\right\}
\eqno (5.8) $$
where $ {\cal B}_U = \bigcup^{ }_{ 0\leq \lambda \leq 1}{\cal
C}^-_\lambda
\times  {\cal C}^+_{1-\lambda} $ and
$$ {\cal C}^+_\mu  = - {\cal C}^-_\mu = \left\{ y : y \in  V_+\cap
\left({\beta \over 2} e + V_- \right),\ \  \left\vert y^2-(e\cdot y)
^2
\right\vert^{ 1/2} \leq  \mu  {\beta \over 4} \right\} \eqno (5.9) $$
for $ 0\leq \mu \leq 1. $}

\vskip 10pt

The proof of this proposition is given in Appendix B. The result can be
compared with the case when $ \omega_ \beta $ is supposed to be strongly
resistant to
boundary effects. There Proposition 4.2 has yielded the tube $ {\cal
T}_{{\cal
B}_M} = - {\cal T}_{{\beta \over 2}e} \times  {\cal T}_{{\beta \over
2}e} $
with basis $ {\cal B}_M = {\cal C}^-_1 \times  {\cal C}^+_1 \supset
{\cal
B}_U $ as a domain of analyticity of $ F. $ (A pictorial
comparison of the bases $ {\cal B}_U $ and $ {\cal B}_M $ is given in
Fig.1.)
Hence in that case the
complete answer (obtained by time-translation invariance of $ F) $ is:

\vskip 12pt

\noindent \underbar{Proposition 5.3} : {\it If the state $ \omega_
\beta $ is
strongly resistant to boundary
effects, then the functions $ F $ of Proposition 5.1, extending the
correlation
functions, are analytic in the tube $ {\cal T}_{{\cal B}} $ with basis
$$ {\cal B }= \left\{ \left(y_1,y_2 \right) : y_1=\eta_ 1+te,\ y_2=\eta_
2+te,\ \left(\eta_ 1,\eta_ 2 \right) \in  {\cal B}_M,\ t \in  \Bbb R
\right\}
\ . \eqno (5.10) $$
}

This is the maximal domain of analyticity of thermal correlation
functions
which one may expect in our general setting.

\midinsert
\vglue 5truecm
\centerline{Fig.1 : The bases $ {\cal B}_U = \bigcup^{ }_{ 0\leq
\lambda \leq
1}{\cal C}^-_\lambda \times{\cal C}^+_{1-\lambda} $ and $ {\cal B}_M =
{\cal
C}^-_1\times{\cal C}^+_1. $}
\endinsert

We now restrict our attention to the physically interesting cases when
the
state
$ \omega_ \beta $ is invariant under the full translation group $ \Bbb
R^d, $
i.e.
$$ \omega_ \beta \left(\alpha_{ x_1}(A)\alpha_{ x_2}(B) \right) =
\omega_
\beta \left(\alpha_{ x_1+x}(A)\alpha_{ x_2+x}(B) \right) \eqno (5.11) $$
for all $ x \in  \Bbb R^d. $ By an argument (based on Lemma A.1)
completely
analogous
to the one previously used for the time-invariance property of $ F, $
one
shows
that the function $ F $ then satisfies the invariance relation
$$ F \left(z_1,z_2 \right) = F \left(z_1+z,\ z_2+z \right)\ ,\ \ \ z
\in  \Bbb
C^d \eqno (5.12) $$
in the convex open tube $ {\cal T}_{{\cal B}} $ where it is defined. It
follows that the resulting
domain of the function $ \underline{F} , $ defined by $ \underline{F}
\left(z_2-z_1 \right) = F \left(z_1,z_2 \right), $ is obtained by
taking the projection of the open tube $ {\cal T}_{{\cal B}} $ in the
space $
\Bbb C^d $ of the vector
variable $ z_2-z_1. $ One can then state:

\vskip 10pt

\noindent \underbar{Corollary 5.4} : {\it Let $ \omega_ \beta $ be a
KMS-state
which is invariant under
space-time translations and satisfies conditions a) and b). There
exists a
function $ \underline{F} , $ analytic in a convex open tube $ {\cal
T}_{{\cal
C}} \subset  \Bbb C^d $ with basis $ {\cal C }\subset  \Bbb R^d, $
which extends the correlation function $ x \longrightarrow  \omega_
\beta
\left(A\alpha_ x(B) \right) $ for given local $ A,B\in{\cal A}. $
More specifically:

\vskip 10pt

i) The basis $ {\cal C} $ is a neighbourhood in $ \Bbb R^d $ of the
linear
segment $ \{ y : y = \lambda e,\ 0< \lambda  < \beta\} , $
and the profile of $ {\cal T}_{{\cal B}} $ at all boundary points in $
\Bbb
R^d $ (resp. in $ \Bbb R^d+i\beta e) $
is the light cone $ V_+ $ (resp. $ \beta e+V_-). $ Moreover, one has
$$ \displaylines{\lim_{ V_+\ni \eta \longrightarrow 0}
\underline{F}(x+i\eta)
= \omega_ \beta \left(A\alpha_ x(B) \right)
\cr\lim_{ V_-\ni \eta
\longrightarrow 0} \underline{F}(x+i\beta e+i\eta)  = \omega_ \beta
\left(\alpha_ x(B)A \right)\ . \cr} $$
A precise shape can be given for the tube $ {\cal T}_{{\cal C}} $ in the
following cases.

\vskip 10pt

ii) If $ \omega_ \beta $ satisfies the condition of resistance to
boundary
effects, then
$$ {\cal C}= \left\{ y : y \in  V_+ \cap
\left(\beta e + V_- \right),\ \
\left\vert y^2-(e\cdot y)^2 \right\vert^{ 1/2} < {\beta \over 4}
\right\} \ .
$$

iii) If $ \omega_ \beta $ satisfies the condition of strong resistance
to boundary effects, then
$$ {\cal C } =  \left\{ y : y \in  V_+ \cap
\left(\beta e
+ V_- \right) \right\} \ , $$
(i.e., ${\cal C}$ is the basis of the tube ${\cal T}_{\beta e}$
introduced in the Introduction).
}

The three typical situations considered in this corollary are depicted
in
Fig.2.

\midinsert
\vglue 5truecm
\centerline{Fig.2 - The basis $ {\cal C} $ in the three statements of
Corollary 5.4.}
\endinsert

The results of this section provide various formulations of the
KMS-condition
which take into account the expected features of thermal correlation
functions in a relativistic theory. Even in their weakest form, namely
Proposition
5.1, they involve all space-time variables and reveal the existence of a
maximal propagation speed through the special role of the light cone $
V_+. $
Moreover, they apply in a canonical way to observers in any Lorentz
frame. We
therefore regard these results as proper versions of a relativistic
KMS-condition.

\vskip 24pt

\noindent {\bf 6. Concluding remarks}

\vskip 12pt

\noindent In the present investigation we have given arguments which
suggest a
relativistic formulation of the KMS-condition for thermal equilibrium
states.
Although we have used the framework of algebraic quantum field theory
for
mathematical convenience, it is apparent that the conclusions of our
analysis
are of a quite general nature and can be applied to unbounded field
operators
(Wightman fields) as well. The resulting analyticity domains of the
Wightman $n$-point functions are of the type described (for $n = 2$) at
the end of the Introduction. These domains are the analogues of the
primitive tube domains implied by the relativistic spectrum condition
in the vacuum case.

As in the latter, these primitive analyticity properties combined
with the condition of locality should define, through the
edge-of-the-wedge technique, the full analytic structure of correlation
functions in complex spacetime. First results in this direction have
been obtained in $\lbrack$BB1$\rbrack$ and $\lbrack$BB2$\rbrack$.

In view of its expected significance for the structural analysis, it
would be
desirable to provide further arguments in favour of a relativistic
version of
the KMS-condition. On the one hand, it would be important to check the
latter
in models. We note in this context that the relativistic KMS-condition
in its
most restrictive form (case iii) of Corollary 5.4) is satisfied by the
thermal equilibrium states of local non-interacting fields and
has also been verified in perturbation theory $\lbrack$St$\rbrack$.
On the other hand, it may well be possible to derive this condition
from more fundamental principles, such as the second law of
thermodynamics.

To motivate the latter statement, we recall that the standard
KMS-condition
can be established for states $ \omega $ which are passive
$\lbrack$PW,BR$\rbrack$, i.e. states from
which one cannot extract energy by a cyclic
process. The condition of passivity is expressed in the mathematical
setting
by the requirement that
$$ \Delta E = i{ {\rm d} \over {\rm d} t} \omega \left. \left(U^\ast
\alpha_{
t\cdot e}(U) \right) \right\vert_{ t=0} \leq  0 \eqno (6.1) $$
for all (differentiable) unitary operators $ U \in  {\cal A}. $ The
quantity $
\Delta E $ can be
interpreted as the energy gained in a cyclic process between the initial
state $ \omega( \cdot) $ and final state $ \omega \left(U^\ast \cdot U
\right)
$ $\lbrack$PW$\rbrack$.

Whereas condition (6.1) is an appropriate expression of the second law
for an
observer in the rest frame of the state $ \omega , $ it is not adequate
for an
observer
who is moving. In fact, the quantity $ \Delta E $ does not take into
account
the energy
which is necessary to maintain the motion of such an observer in the
presence
of dissipative forces; as a result $ \Delta E $ can become strictly
positive.

It is clear, however, that the energy $ \Delta E $ has to be smaller
than the
energy
fed into the system by the moving observer. This fact suggests to amend
the
passivity condition (6.1) by the following assumption: for any positive
timelike vector $ f \in  V_+, $ $ f^2=1, $ and any bounded spacetime
region $
{\cal O} $ there exists
a constant $ E_{f,{\cal O}} $ such that
$$ i { {\rm d} \over {\rm d} t} \omega \left. \left(U^\ast \alpha_{
t\cdot
f}(U) \right) \right\vert_{ t=0}\leq  E_{f,{\cal O}} \eqno (6.2) $$
for all (differentiable) unitary operators $ U \in  {\cal A}({\cal O}).
$ The
essence of this
condition is the assumption that the energy $ E_{f,{\cal O}} $ which is
necessary to
proceed from the rest system to the moving system, characterized by $
f, $ is
locally finite. The dependence of $ E_{f,{\cal O}} $ on $ f $ will be
submitted to the specific
properties of the state $ \omega , $ but one may expect that, quite
generally,
$ E_{f,{\cal O}} $ is
proportional to the size of $ {\cal O} $ for large spacetime regions $
{\cal
O}. $

It seems worthwhile to explore the consequences of condition (6.2) for
the
structure of the correlation functions. We hope that by applying the
powerful
methods developped in $\lbrack$PW$\rbrack$ it will be possible to
establish
analyticity
properties of these functions, as anticipated in our stability
condition of
Sec.\ 5. This would then provide an alternative and quite fundamental
justification of our relativistic version of the KMS-condition.

\vskip 24pt

\vfill\eject
\noindent {\bf Appendix A}

\vskip 12pt

\noindent We collect here some definitions and results of the theory of
analytic functions of several complex variables which are used
throughout Sec.\ 5.

The functions $ F $ which we encounter are typically defined in subsets
of $
\Bbb C^N $
which are tubes, namely sets of the form $ {\cal T}_{{\cal B}} = \Bbb
R^N+i{\cal B}, $ the set $ {\cal B }\subset  \Bbb R^N $
being called the basis of the tube $ {\cal T}_{{\cal B}}. $ If $ {\cal
B} $ is
an open set in $ \Bbb R^N, $ $ {\cal T}_{{\cal B}} $ is
an open tube in $ \Bbb C^N; $ the analyticity of a function $ F $ in $
{\cal
T}_{{\cal B}} $ then means
analyticity of $ F(z) $ with respect to all complex variables $ z =
\left(z_1,...,z_N \right) $
varying in $ {\cal T}_{{\cal B}}. $

If $ {\cal B} $ is a linear submanifold of $ \Bbb R^N $ of dimension $
n < N,
$ $ {\cal T}_{{\cal B}} $ will be called
a {\sl flat tube\/}; by analyticity of a function $ F $ in the flat
tube $
{\cal T}_{{\cal B}}, $ we shall
always mean: joint continuity with respect to all variables varying in
$ {\cal
T}_{{\cal B}} $ and
analyticity of $ F $ with respect to a (maximal) set of $ n $ complex
variables in
all the complex $ n $-dimensional (linear) submanifolds which generate
$ {\cal
T}_{{\cal B}}. $
A simple but important example of a flat tube in $ \Bbb C^2 $ is $ {\cal
T}_{{\cal B}}=\Bbb R^2+i{\cal B}, $
where $ {\cal B }= \left\{ \left(y_1,y_2 \right)\in \Bbb R^2 : \right.
$ $ 0 <
y_1 < b, $ $ \left.y_2 = 0 \right\} . $

We shall also be led to consider tubes $ {\cal T}_{\cal B}$ with a
non-empty interior which are
obtained by adjoining one or several flat tubes to their interior (in
other
words, the basis ${\cal B}$ of such a tube is the union of an open
connected set and
of one or several linear manifolds belonging to the boundary of the
latter):
such tubes may be produced by taking the convex hull of the union of
two or
several flat tubes (see Lemma A.2 below and in particular the first case
presented in the proof of the latter). In such a case, we shall say
that $
{\cal T}_{\cal B}$
is a \lq\lq flatly-bordered tube\rq\rq\ and that a function $ F $ is
analytic
in $ {\cal T}_{\cal B}$ if the
following conditions are fulfilled: i) $ F $ is continuous in
$ {\cal T}_{\cal B}, $
ii) the restriction of $ F $ to the interior of $ {\cal T}_{\cal B}$ is
analytic; this
implies that the
restrictions of $ F $ to the various bordering flat tubes in
$ {\cal T}_{\cal B}$ are also analytic
in the sense of the previous definition.

If a point $ b $ of $ \Bbb R^N $ belongs to the boundary of
${\cal B}, $ we call
{\sl profile\/} of the
tube $ {\cal T}_{{\cal B}} $ at any point $ c = a+ib $ of its boundary
the
cone $ \Lambda_ b $ with apex $ b $ in $ \Bbb R^N $
which is the union of all closed half-lines starting from $ b $ and
intersecting $ {\cal B}. $ We will say that a function $ F, $ analytic
in $
{\cal T}_{{\cal B}}, $ admits a
{\sl continuous boundary value\/} near a boundary point $ c = a+ib $ of
$
{\cal T}_{{\cal B}} $ if it can be
extended as a continuous function on some neighbourhood of $ c $ in $
{\cal
T}_{{\cal B}\cup\{ b\}} $ in the
following sense: for each closed subcone $ \bar \Lambda $ of $ \Lambda_
b $
there exists a real
neighbourhood $ {\cal N}_a $ of $ a $ such that the extension of $ F $
is
continuous in the
region $ {\cal N}_a+i\bar \Lambda . $ Similarly, $ F $ is said to be
continuous on a given open set of
boundary points if it is continuous in the above sense near each of
these
points.

The following result can be seen as an extension of the principle of
uniqueness of the analytic continuation (see e.g. $\lbrack$SW$\rbrack$
Theorem
2.17).

\vskip 12pt

\noindent \underbar{Lemma A.1} : {\it Let $ {\cal T}_{{\cal B}} $ be a
given
(open or flat) tube and $ c = a+ib $ a
boundary point of $ {\cal T}_{{\cal B}}. $ If two functions $ F $ and $
G, $
analytic in $ {\cal T}_{{\cal B}}, $ admit
coinciding (continuous) boundary values near the point $ c, $ then $
F=G. $}

\vskip 10pt

The basic result concerning analytic functions in tube-shaped domains
is the
following {\sl tube theorem\/} (see e.g. $\lbrack$W$\rbrack$ and
references
therein): if a function $ F $
is analytic in a tube $ {\cal T}_{{\cal B}} $ with open connected basis
$
{\cal B}, $ then it can be
analytically continued in the tube $ {\cal T}_{\widehat{\cal B}} $
whose basis $ \widehat{\cal B } $
                                   is the convex hull of $ {\cal B}. $
In other words: $ {\cal T}_{\widehat{\cal B }} $ is the holomorphy
envelope of $ {\cal T}_{{\cal B}}. $

A non-trivial refinement of the tube theorem is the fact that it can be
extended to the case of flat tubes (see e.g. $\lbrack$BEGS$\rbrack$,
the first
result of this
type being due to Malgrange and Zerner). We shall need the following
version
of this {\sl flat tube theorem\/}.

\vskip 12pt

\noindent \underbar{Lemma A.2} : {\it Let $ {\cal T}_{{\cal B}_0} $ and
$
{\cal T}_{{\cal B}_1} $ be two flat tubes in $ \Bbb C^N $ whose
bases $ {\cal B}_i, $ $ i=0,1 $ are convex and have closures
$ \bar {\cal B}_i $
which contain the origin $ 0 $
and are star-shaped with respect to $ 0. $ Let $ F_0 $ and $ F_1 $ be
any pair
of functions
which are analytic in $ {\cal T}_{{\cal B}_0} $ and $ {\cal T}_{{\cal
B}_1} $
respectively, and have continuous boundary
values on $ \Bbb R^N $ which coincide, i.e. $ F_0 \upharpoonright \Bbb
R^N =
F_1 \upharpoonright \Bbb R^N. $
Then there exists a unique function $ F $ which is analytic in the tube
$
{\cal T}_{ \widehat{{\cal B}_0\cup{\cal B}_1}}, $
has continuous boundary values on $ \Bbb R^N, $ and extends the given
functions:
$$ F \upharpoonright {\cal T}_{{\cal B}_i} = F_i,\ \ \ \ \ i = 0,1
\ . $$
Moreover, the convex tube $ {\cal T}_{ \widehat{{\cal B}_0\cup{\cal
 B}_1}} $
can be described as follows:
$$ {\cal T}_{ \widehat{{\cal B}_0\cup{\cal B}_1}} = \bigcup^{ }_{ 0\leq
\lambda \leq 1}{\cal T}_{{\cal B}_\lambda} \ , $$
where
$$ {\cal B}_\lambda  = (1-\lambda){\cal B}_0+\lambda{\cal B}_1 =
\left\{ y\in
\Bbb R^N : y = (1-\lambda) y^{(0)}+\lambda y^{(1)},\ y^{(0)}\in{\cal
B}_0,\
y^{(1)}\in{\cal B}_1 \right\} \ . $$
}

\vskip 10pt

We notice that if $ {\cal B}_0 $ and $ {\cal B}_1 $ are linearly
independent
and of respective
dimensions $ n_0 $ and $ n_1, $ $ \widehat{{\cal B}_0\cup{\cal B}_1} =
\bigcup^{ }_{ 0\leq \lambda \leq 1}{\cal B}_\lambda $ is of dimension $
n_0+n_1. $ Thus the
non-trivial aspect of this result is that analyticity with respect to $
n_0+n_1 $
variables is obtained from two assumptions of analyticity with respect
to $
n_0 $
and $ n_1 $ variables in the distinct sets $ {\cal T}_{{\cal B}_0} $
and $
{\cal T}_{{\cal B}_1}. $ Of course, the coincidence
condition on the reals is crucial and plays the role of the
connectedness of
the tube $ {\cal T}_{{\cal B}} $ in the standard tube theorem. We also
notice
that the uniqueness of the
function $ F $ is a direct consequence of the analytic continuation
principle
(respectively of Lemma A.1).

\vskip 12pt

\noindent \underbar{Proof of Lemma A.2} : Let us first consider a
simple case
covered by
the Lemma which is in fact basic for the proof of the general case, as
indicated below. This is the two-dimensional situation where $ {\cal
T}_{{\cal
B}_i} \subset  \Bbb C^2, $ $ i=0,1 $
are tubes with bases $ {\cal B}_0 = \left\{ \left(y_1,y_2 \right) :
\right. $
$ 0 < y _1 < a_1, $ $ \left.y_2 = 0 \right\} $ and $ {\cal B}_1 =
\left\{
\left(y_1,y_2 \right) : \right. $
$ y_1=0, $ $ \left.0< y_2 < a_2 \right\} . $ The corresponding convex
hull is
then
$$ \widehat{{\cal B}_0\cup{\cal B}_1} = \left\{ \left(y_1,y_2 \right) :
0 \leq
y_1,\ 0\leq  y_2,\ 0 < {y_1 \over a_1} + {y_2 \over a_2} < 1 \right\} \
 , $$
equivalently described as
$$ \widehat{{\cal B}_0\cup{\cal B}_1} = \left( \bigcup^{ }_{ 0<\lambda
<1}
\left\{ \left(y_1,y_2 \right) : 0 < y_1 < (1-\lambda) a_1,\ 0 < y_2 <
\lambda
a_2 \right\} \right) \bigcup^{ }_{ } \left({\cal B}_0\cup{\cal B}_1
\right) $$

The proof of Lemma A.2 in this case is essentially given in
$\lbrack$BEGS$\rbrack$. In the
version presented there, $ F_0 $ and $ F_1 $ are assumed to be $
C^{\infty} $
functions\footnote{$ ^{7)} $
}{{\sevenrm In $\lbrack$BEGS$\rbrack$, a direct proof of the flat tube
property, based on a Cauchy
integral method, is first given for the case when $ F_0 $ and $ F_1 $
are $
C^{\infty} $ and
sufficiently decreasing at infinity (Lemma 1); a {\sl localized\/}
version of
the
flat tube property is then derived from the latter by an appropriate
use of
conformal mappings (Lemma 2); finally, as indicated in a subsequent
remark, a
by-product of this localized version is that it allows one to get rid
of any
restriction on the behaviour at infinity of $ F_0, $ $ F_1 $ in the
first
result; it
actually displays the purely local character of the analytic completion
procedure which is at work in the (flat) tube property.}}.

However, the present version (in which $ F_0 $ and $ F_1 $ are only
assumed to
be
continuous) can be easily traced back to the situation considered in
$\lbrack$BEGS$\rbrack$
by approximating the given functions $ F_0, $ $ F_1 $ by sequences of $
C^{\infty} $-functions $ F^{(n)}_i=F_i\ast \delta_ n, $
$ i=0,1, $ where $ \left\{ \delta_ n,\ n\in \Bbb N \right\} $ is a
regularizing sequence of test functions in $ {\cal D} \left(\Bbb R^2
\right) $
which tends to the Dirac measure in the limit of large $ n. $ In view
of the
results of $\lbrack$BEGS$\rbrack$ (and of the Cauchy integral method
used
therein$ ^{7)}), $ the
smooth functions $ F^{(n)}_i, $ $ i=0,1, $ can be continued to analytic
functions $ F^{(n)} $ in
$ {\cal T}_{ \widehat{{\cal B}_0\cup{\cal B}_1}} $ which (by virtue of
the
maximum modulus principle) are uniformly
bounded on all compact subsets of $ {\cal T}_{ \widehat{{\cal
B}_0\cup{\cal
B}_1}}\cup \Bbb R^2. $ It follows that
the functions $ F^{(n)} $ form a normal family of analytic functions,
and that
the
desired analytic continuation $ F $ of the functions $ F_0, $ $ F_1 $
is then
defined as a
limit point of this normal family.

The proof of the multi-dimensional version of Lemma A.2 relies on the
two-dimensional result thanks to the following simple geometrical
argument.
One sweeps each of the (star-shaped) bases $ {\cal B}_0, $ $ {\cal B}_1
$ of
the tubes $ {\cal T}_{{\cal B}_0}, $ $ {\cal T}_{{\cal B}_1} $ by
linear segments $ {\cal L}_0 = \left\{ y : y = \lambda \cdot b_0,
\right. $ $
0 < \lambda  < 1\} $ for any $ b_0 \in  {\cal B}_0, $ and $ {\cal L}_1 =
\left\{ y : y=\lambda \cdot b_1, \right. $
$ 0 < \lambda  < 1\} $ for any $ b_1 \in  {\cal B}_1. $ By applying the
two-dimensional flat tube
theorem, quoted above, to the couple of functions $ F_0 \upharpoonright
{\cal
T}_{{\cal L}_0} $ and
$ F_1 \upharpoonright {\cal T}_{{\cal L}_1} $ one obtains a
corresponding
analytic continuation $ F_{b_0b_1} $
in the flat convex tube $ {\cal T}_{ \widehat{{\cal L}_0\cup{\cal
L}_1}} =
\bigcup^{ }_{ 0\leq \lambda \leq 1}{\cal T}_{(1-\lambda){\cal
L}_0+\lambda{\cal L}_1}. $ As a matter of fact, the
germs of functions $ F_{b_0b_1} $ obtained in each of these flat tubes
are
analytic
not only with respect to the two complex variables associated with the
complex two-plane fixed by $ b_0, $ $ b_1, $ but with respect to all
those
variables
which vary in the full tube $ {\cal T}_{ \widehat{{\cal B}_0\cup{\cal
B}_1}}.
$ This can be seen directly by inspection
of the proof of Lemma 1 in $\lbrack$BEGS$\rbrack$, where the Cauchy
integrals
used for
defining $ F_{b_0b_1} $ now exhibit (through $ F_0, $ $ F_1) $ an
analytic
dependence with
respect to all variables involved in $ {\cal T}_{{\cal B}_0} $ and $
{\cal
T}_{{\cal B}_1}. $ Finally, patching together
all these germs of analytic functions $ F_{b_0b_1} $ for $ b_0 \in
{\cal B}_0
$ and $ b_1\in{\cal B}_1 $ yields a
univalent analytic function $ F $ in the (simply connected) domain $
\bigcup^{
}_{ 0\leq \lambda \leq 1}{\cal T}_{{\cal B}_\lambda} , $ since
$ {\cal B}_\lambda  = \bigcup^{ }_{ b_0\in{\cal B}_0,b_1\in{\cal B}_1}
\left((1-\lambda)  {\cal L}_0+\lambda{\cal L}_1 \right). $ The fact
that $
\bigcup^{ }_{ 0\leq \lambda \leq 1}{\cal B}_\lambda $ is convex and
therefore equal to the convex hull of $ {\cal B}_0\cup{\cal B}_1 $ can
be checked directly.

\vskip 10pt

We shall also make use of a variant of the tube and flat tube theorems,
in
which the basis $ {\cal B} $ of the tube $ {\cal T}_{{\cal B}} $ is
\lq\lq{\sl dumb-bell shaped\/}\rq\rq , i.e. of the form $ {\cal B}={\cal
B}_0\cup{\cal B}_1\cup{\cal L}, $
where $ {\cal B}_0 $ and $ {\cal B}_1 $ are two disjoint convex open
sets of $
\Bbb R^N $ and $ {\cal L} $ is a
linear segment whose end-points $ b_0, $ $ b_1 $ belong respectively to
$
{\cal B}_0 $ and $ {\cal B}_1. $ A
function $ F $ is said to be analytic in the tube $ {\cal T}_{{\cal B}}
$ with
dumb-bell shaped basis
$ {\cal B}={\cal B}_0\cup{\cal B}_1\cup{\cal L} $ if it is continuous
on $
{\cal T}_{{\cal B}}, $ analytic in $ {\cal T}_{{\cal B}_0} $ and $ {\cal
T}_{{\cal B}_1} $ (as a function
of $ N $ complex variables) and analytic in $ {\cal T}_{{\cal L}} $ (as
a
function of one complex
variable).

\vskip 12pt

\noindent \underbar{Lemma A.3} : {\it Any function $ F $ which is
analytic in
a tube $ {\cal T}_{{\cal B}} $ with
dumb-bell shaped basis $ {\cal B}={\cal B}_0\cup{\cal B}_1\cup{\cal L}
$ can
be analytically continued (as a function
of $ N $ complex variables) in the tube
$ {\cal T}_{\widehat{\cal B }} $ whose
basis $ \widehat{\cal B} $ is the convex hull of
$ {\cal B}_0\cup{\cal B}_1. $}

\vskip 10pt

\noindent \underbar{Proof} : Without restriction of generality we may
assume
that the
endpoint $ b_0 $ of $ {\cal L} $ is the origin $ 0 $ of $ \Bbb R^N, $
i.e.
that $ 0 \in  {\cal B}_0. $ Let then $ a_1,...,a_N $
be $ N $ points in $ {\cal B}_0 $ whose convex hull $ {\cal H} $ is an
$ (N-1)
$-dimensional simplex
containing $ 0 $ as an interior point, and let us consider the two flat
tubes
$ {\cal T}_{{\cal H}} $
and $ {\cal T}_{{\cal L}}. $ It is clear that any function $ F, $
analytic in
$ {\cal T}_{{\cal B}}, $ defines a pair of
functions $ F_0 = F \upharpoonright {\cal T}_{{\cal H}} $ and $ F_1 = F
\upharpoonright {\cal T}_{{\cal L}} $ which satisfy
all the conditions of Lemma A.2. As a result, there exists an analytic
function $ F_{01} $ of $ N $ variables which is the common analytic
continuation of $ F_0 $
and $ F_1 $ in the convex tube $ {\cal T}_{ \widehat{{\cal H}\cup{\cal
L}}} =
\bigcup^{ }_{ 0\leq \lambda \leq 1}{\cal T}_{(1-\lambda){\cal
H}+\lambda{\cal
L}}. $ Since $ F_{01} \upharpoonright \Bbb R^N = F \upharpoonright \Bbb
R^N, $
it follows from Lemma A.1 that the restrictions of $ F_{01} $ and $ F $
to the
domain $ {\cal T}_{{\cal B}_0}\cap{\cal T}_{ \widehat{{\cal H}\cup{\cal
L}}} $
coincide; similarly, since $ F_{01} \upharpoonright \left(\Bbb R^N+ib_1
\right) = F \upharpoonright \left(\Bbb R^N+ib_1 \right), $
it follows that the restrictions of $ F_{01} $ and $ F $ to the domain
$ {\cal
T}_{{\cal B}_1}\cap{\cal T}_{ \widehat{{\cal H}\cup{\cal L}}} $
coincide. The function $ F_{01} $ therefore provides an analytic
continuation
of $ F $
in the connected open tube $ {\cal T}_{{\cal B}_0} \cup  {\cal
T}_{{\cal B}_1}
\cup  {\cal T}_{ \widehat{{\cal H}\cup{\cal L}}}. $ Hence, by applying
the
standard
tube theorem, we conclude that $ F $ can be analytically continued in
the tube
$ {\cal T}_{\widehat{\cal B}}, $
whose basis $ \widehat{\cal B}  $ is the convex hull of
$ {\cal B}_0 \cup {\cal
B}_1 \cup  \left( \widehat{{\cal H}\cup{\cal L}} \right) $ and therefore
coincides
with the convex hull of $ {\cal B}_0 \cup  {\cal B}_1. $

\vskip 24pt

\noindent {\bf Appendix B}

\vskip 12pt

\noindent We give here the proof of Proposition 5.2. The argument is
similar to the one given in
Proposition 5.1, but the more detailed statement about the shape of the
tube
$ {\cal T}_{{\cal B}} $ requires some extra calculations.

From the KMS-condition a) and the assumption that $ \omega_ \beta $ is
resistant to boundary
effects, we obtain analyticity of $ F $ (together with continuity at the
edges)
in the four flat tubes $ {\cal T}_{{\cal B}_1} = \Bbb R^d \times
{\cal T}_{{\beta \over 2}e}, $ $ {\cal T}_{{\cal B}^{\dagger}_ 1}=
\left(-{\cal T}_{{\beta \over 2}e}
\right)\times  \Bbb R^d, $ $ {\cal T}_{{\cal B}_2}= \left(\Bbb R^d -i
{\beta
\over 2}e \right)\times  {\cal T}_{{\beta \over 2}e} $
and $ {\cal T}_{{\cal B}^{\dagger}_ 2}= \left(-{\cal T}_{{\beta \over
2}e}
\right)\times  \left(\Bbb R^d+i {\beta \over 2}e \right). $ Lemma A.2
can now
be applied to the four
pairs $ \left({\cal T}_{{\cal B}_1},{\cal T}_{{\cal B}^{\dagger}_ 1}
\right),
$ $ \left({\cal T}_{{\cal B}_2},{\cal T}_{{\cal B}^{\dagger}_ 2}
\right), $ $
\left({\cal T}_{{\cal B}_1},{\cal T}_{{\cal B}^{\dagger}_ 2} \right) $
and $
\left({\cal T}_{{\cal B}_2},{\cal T}_{{\cal B}^{\dagger}_ 1} \right). $
Since
the four resulting
convex tubes are flatly bordered tubes in $ \Bbb C^d \times  \Bbb C^d,
$ a
final application of the
standard tube theorem to the union of the interiors of the latter yields
analyticity of $ F $ in the convex tube $ {\cal T}_{{\cal B}}, $ whose
basis $ {\cal B} $ is the convex hull of $ {\cal B}_1 \cup  {\cal
B}^{\dagger}_ 1 \cup  {\cal B}_2 \cup  {\cal B}^{\dagger}_ 2. $ (The
uniqueness of $ F, $ continued
into the various common domains, is ensured by Lemma A.1.)

We are now just led to the technical problem of computing the convex
hull $
{\cal B}, $
which we treat as follows: let $ {\cal L}^{\dagger}_ 1 = \left\{
\left(y_1,y_2
\right) : y_1=\lambda e, \right. $ $ -\beta /2 < \lambda  < 0, $ $
\left.y_2=0
\right\} $ be
the \lq\lq diagonal\rq\rq\ of the base $ {\cal B}^{\dagger}_ 1. $ The
convex
hull of $ {\cal B}_1\cup{\cal L}^{\dagger}_ 1\cup{\cal B}_2 $ is the $
(d+1 $
-dimensional) set $ {\cal C}^-_0\times{\cal C}^+_1, $ where
(consistently with
the notations used in the
statement of the proposition)
$$ {\cal C}^-_0 = \left\{ y:y=\lambda e,\ -{\beta \over 2} < \lambda  <
0
\right\} \ \ {\rm and} \ \ \ {\cal C}^+_1 = \left\{ y:y\in V^+\cap
\left({\beta \over 2}e+V^- \right) \right\} \ . $$
Similarly, if $ {\cal L}_1= \left\{ \left(y_1,y_2 \right):y_1=0,
\right. $ $
y_2=\lambda \cdot e, $ $ \left.0< \lambda  < {\beta \over 2} \right\} $
is the
diagonal of $ {\cal B}_1, $
the convex hull of $ {\cal B}^{\dagger}_ 1 \cup  {\cal L}_1 \cup  {\cal
B}^{\dagger}_ 2 $ is the set $ {\cal C}^-_1\times{\cal C}^+_0, $ where
$$ {\cal C}^-_1= \left\{ y:y\in V^-\cap \left(-{\beta \over 2}e +V^+
\right)
\right\} \ \ \ {\rm and} \ \ \ {\cal C}^+_0 = \left\{ y:y=\lambda e,\ 0<
\lambda  < {\beta \over 2} \right\} \ . $$
It is now straightforward to compute $ {\cal B} $ by noting that it is
the
convex hull
of the region $ \left({\cal C}^-_0\times{\cal C}^+_1 \right)\cup
\left({\cal
C}^-_1\times{\cal C}^+_0 \right). $ Alternatively, $ {\cal B} $ can be
characterized as being
the union of all interpolating products $ {\cal C}^-_\lambda  \times
{\cal
C}^+_{1-\lambda} $ for $ 0 \leq  \lambda  \leq  1. $ The latter
fact can be seen by taking arbitrary two-dimensional meridian sections
of the
double cones $ {\cal C}^-_1 $ and $ {\cal C}^+_1. $ One thereby obtains
products of interpolating
trapezia for the corresponding convex completions in the chosen
products of
meridian sections, as indicated in Fig.3.

\midinsert
\vglue 8truecm
\centerline{Fig.3 : Interpolating trapezia}

a)\nobreak\ a meridian section of $ {\cal C}^-_1 $ (represented with the
origin at $ -{\beta \over 2}e) $ with
hatchings inside $ {\cal C}^-_\lambda $
b)\nobreak\ a meridian section of $ {\cal C}^+_1 $ with hatchings
inside $
{\cal C}^+_{1-\lambda} . $
\endinsert

By taking finally into account the time invariance of $ F, $ as in the
proof
of
Proposition 5.1, the statement then follows.

\vfill\eject

\noindent {\bf Acknowledgements}

\vskip 24pt

\noindent The authors are grateful for financial support and
hospitality granted to them repectively (for J.\ B.) by the
franco-german science cooperation PROCOPE and the II. Institut f\"ur
Theoretische Physik, Universit\"at Hamburg , and (for D.\ B.)
by the Service de Physique Th\'eorique de Saclay, CEA.

\vfill\eject

\centerline{{\bf REFERENCES}}

\vglue 1truecm

\item{$\lbrack$BR$\rbrack$}Bratteli, O., Robinson, D.W.: Operator
algebras and
quantum
statistical mechanics II. Berlin, Heidelberg, New York: Springer 1981.

\item{$\lbrack$H$\rbrack$}Haag, R.: Local quantum physics, Berlin,
Heidelberg,
New York:
Springer 1992.

\item{$\lbrack$HHW$\rbrack$}Haag, R., Hugenholtz, N.M. and Winnink, M.:
On the
equilibrium
states in quantum statistical mechanics, {\sl Commun. Math. Phys.\/}
{\bf 5},
215-236 (1967).

\item{$\lbrack$HKTP$\rbrack$}Haag, R., Kastler, D. and Trych-Pohlmeyer,
E.:
Stability and
equilibrium states, {\sl Commun. Math. Phys.\/} {\bf 38}, 173-193
(1974).

\item{$\lbrack$PW$\rbrack$}Pusz, W., Woronowicz, S.L.: Passive states
and
KMS-states for
general quantum systems, {\sl Commun. Math. Phys.\/} {\bf 58}, 273-290
(1978).

\item{$\lbrack$O$\rbrack$}Ojima, I.: Lorentz invariance vs. temperature
in
QFT, {\sl Lett. Math.
Phys.\/} {\bf 11}, 73-80\nobreak\ \nobreak\ \nobreak\  (1986).

\item{$\lbrack$N$\rbrack$}Narnhofer, H.: Kommutative Automorphismen und
Gleichgewichtszust\"ande,{\sl\ Act. Phys. Austriaca\/} {\bf 47}, 1-29
(1977).

\item{$\lbrack$D$\rbrack$}Dixmier, J.: Von Neumann algebras, Amsterdam,
New
York, Oxford:
North Holland 1981.

\item{$\lbrack$BW$\rbrack$}Buchholz, D., Wichmann, E.H.: Causal
independence
and the
energy-level density of states in local quantum field theory, {\sl
Commun.
Math.
Phys.\/} {\bf 106}, 321-344 (1986).

\item{$\lbrack$BJ$\rbrack$}Buchholz, D., Junglas, P.: On the existence
of
equilibrium states
in local quantum field theory, {\sl Commun. Math. Phys.\/} {\bf 121},
255-270
(1989).

\item{$\lbrack$K$\rbrack$}Kato, T.: Perturbation theory for linear
operators,
New York:
Springer 1966.

\item{$\lbrack$Sch$\rbrack$}Schlieder, S.: Einige Bemerkungen \"uber
Projektionsoperatoren,
{\sl Commun. Math. Phys.\/} {\bf 13}, 216-225 (1969).

\item{$\lbrack$S$\rbrack$}Sakai, S.: $ C^\ast $-algebras and $ W^\ast
$-algebras, Berlin, Heidelberg, New
York: Springer 1971.

\item{$\lbrack$BEGS$\rbrack$}Bros, J., Epstein, H., Glaser, V. and
Stora, R.:
Quelques
aspects globaux des probl\`emes d'edge of the wedge, pp.185-218 in:
Hyperfunctions and theoretical physics. Lecture Notes in Mathematics
449,
Berlin, Heidelberg, New York: Springer 1975.

\item{$\lbrack$BB1$\rbrack$}Bros, J., Buchholz, D.: Particles and
propagators
in relativistic
thermo field theory, {\sl Z. Phys. C\/} - Particles and Fields {\bf 55},
509-513 (1992).

\item{$\lbrack$BB2$\rbrack$}Bros, J., Buchholz, D.: Fields at finite
temperature: A general
study of the two-point function, to appear.

\item{$\lbrack$St$\rbrack$}Steinmann, O.: Private communication. Cf.
also: Perturbative quantum field theory at positive temperatures.
An axiomatic approach, Preprint Universit\"at Bielefeld (1994)

\item{$\lbrack$SW$\rbrack$}Streater, R.F., Wightman, A.S., PCT, spin and
statistics, and all
that, New York, Amsterdam: Benjamin 1964.

\item{$\lbrack$W$\rbrack$}Wightman, A.S.: Analytic functions of several
complex variables,
pp.159-221 in: Dispersion relations and elementary particles, New York:
Wiley
1960.
\end